\documentclass[prd,aps,twocolumn,preprintnumbers, showpacs, nofootinbib,superscriptaddress,notitlepage]{revtex4-2}
\usepackage{mathrsfs}
\usepackage{amsfonts}
\usepackage{amsmath}
\usepackage{slashed}
\usepackage{array}
\usepackage{verbatim}
\usepackage{epsfig}
\usepackage{graphicx}
\usepackage{color}
\usepackage{hyperref}
\usepackage[dvipsnames]{xcolor}
\usepackage{subfigure}
\usepackage{float}

\newcommand{\beq}{\begin{eqnarray}}
\newcommand{\eeq}{\end{eqnarray}}

\def\be{\begin{equation}}
\def\ee{\end{equation}}
\def\bea{\begin{eqnarray}}
\def\eea{\end{eqnarray}}

\begin{document}

\title{Resumming Quark's Longitudinal Momentum Logarithms \\in LaMET Expansion of 
Lattice PDFs}

\author{Yushan Su}
\email{ysu12345@umd.edu}
\affiliation{Department of Physics, University of Maryland, College Park, Maryland 20742, USA} 
\affiliation{Physics Division, Argonne National Laboratory, Lemont, Illinois 60439, USA}

\author{Jack Holligan}
\email{jeholligan@gmail.com}
\affiliation{Department of Physics, University of Maryland, College Park, Maryland 20742, USA}
\affiliation{Center for Frontier Nuclear Science, Stony Brook University, Stony Brook, NY 11794, USA}

\author{Xiangdong Ji}
\email{xji@umd.edu}
\affiliation{Department of Physics, University of Maryland, College Park, Maryland 20742, USA} 

\author{Fei Yao}
\email{feiyao@mail.bnu.edu.cn}
\affiliation{Center of Advanced Quantum Studies, Department of Physics, Beijing Normal University, Beijing 100875, China}

\author{Jian-Hui Zhang}
\email{zhangjianhui@cuhk.edu.cn}
\affiliation{School of Science and Engineering, The Chinese University of Hong Kong, Shenzhen 518172, China}
\affiliation{Center of Advanced Quantum Studies, Department of Physics, Beijing Normal University, Beijing 100875, China}

\author{Rui Zhang}
\email{rayzhang@umd.edu}
\affiliation{Department of Physics, University of Maryland, College Park, Maryland 20742, USA}

\date{\today}

\begin{abstract}

In the large-momentum expansion for parton distribution functions (PDFs), 
the natural physics scale is the longitudinal 
momentum ($p_z$) of the quarks (or gluons) in a large-momentum hadron. 
We show how to expose this scale dependence through resumming
logarithms of the type $\ln^n p_z/\mu$ in the matching
coefficient, where $\mu$ is a fixed renormalization scale. The result enhances the accuracy of the expansion at moderate $p_z>1$~GeV, and at the same time, 
clearly shows that the partons cannot
be approximated from quarks with $p_z \sim \Lambda_{\rm QCD}$ which are not predominantly
collinear with the parent hadron momentum, consistent with
power counting of the large-momentum effective theory. The same
physics mechanism constrains the coordinate 
space expansion at large distances $z$, the conjugate of $p_z$, 
as illustrated in the example of fitting the moments of the PDFs.

\end{abstract}

\maketitle

\section{Introduction}\label{sec:intro}

Large-momentum effective theory (LaMET) is an approach to calculate
parton physics through large-momentum expansion of Euclidean observables, such 
as momentum distributions and static correlations
calculable in lattice quantum chromodynamics (QCD)~\cite{Ji:2013dva,Ji:2014hxa}. 
It has a wide range of applications, including quark isovector distribution functions~\cite{Lin:2014zya,Alexandrou:2015rja,Chen:2016utp,Alexandrou:2016jqi,Alexandrou:2018pbm,Chen:2018xof,Lin:2018pvv,Liu:2018uuj,Alexandrou:2018eet,Liu:2018hxv,Chen:2018fwa,Izubuchi:2019lyk,Shugert:2020tgq,Chai:2020nxw,Lin:2020ssv,Fan:2020nzz,Gao:2021dbh,Gao:2022iex}, generalized parton distributions~\cite{Chen:2019lcm,Alexandrou:2019dax,Lin:2020rxa,Alexandrou:2020zbe,Lin:2021brq,Scapellato:2022mai}, distribution amplitudes (DAs)~\cite{Zhang:2017bzy,Chen:2017gck,Zhang:2020gaj,Hua:2020gnw, Hua:2022kcm}, and transverse-momentum-dependent distributions~\cite{Shanahan:2019zcq,Shanahan:2020zxr,Zhang:2020dbb,LPC:2022ibr}. Recent reviews on LaMET can be found in Refs.~\cite{Cichy:2018mum,Ji:2020ect}.

For collinear PDFs, the LaMET expansion starts from a quasi-PDF matrix element, which is motivated from the spatial correlator defining the momentum distribution in a many-body system:
\bea\label{eq:quasi}
\tilde{h}^{\rm lat}(z, a, P_{z}) = \langle P_z| O_{\Gamma}(z)|P_z \rangle,
\eea
where $|P_z\rangle$ is the hadron state with a large momentum $P_z$ along $z$ direction.
 $O_{\Gamma}(z)$ is the quasi-PDF operator:
\bea\label{eq:operator}
O_{\Gamma}(z)=\bar{\psi}(z) \Gamma U(z,0) \psi (0),
\eea
where $\psi$, $\bar{\psi}$ denote the bare quark field, $\Gamma$ is a Dirac structure,  and $U(z, 0) = P\exp(-ig\int_0^{z} dz' A_z(z'))$ is the Wilson gauge-link along the direction $z$, where $A^\mu$ is the gluon gauge potential. Apart from the correlator above, there are other choices of Euclidean correlators that can be used to probe the same lightcone physics in the PDFs through large-momentum expansion~\cite{Ji:2020ect}. They form a universality class.

In the large momentum limit, the quasi-PDF or longitudinal momentum distribution can be converted to the standard PDF through a perturbative matching~\cite{Ji:2020ect}. Such a matching involves logarithms associated with the hadron momentum $P_z$ and the renormalization scale $\mu$. Since $P_z$ and $\mu$ are independent scales, their ratio can be large, generating large logarithms that need to be resummed to obtain a more reliable result. It is the purpose of this work to study the effects of these potentially large logarithms and their resummation through standard renormalization group equation (renormalization group resummation or RGR). One may wonder whether resummation is necessary in present lattice calculations where $P_z\sim \mu\sim 2$~GeV. However, as we will argue, the physical scale associated with the quasi-PDF is actually the quark
or gluon longitudinal momentum, $p_z=xP_z$ with $x$ being the momentum fraction. 
It can be much smaller than $\mu$ if $x$ is small. After resummation, the intrinsic scale 
$\sim x P_z$ appears in the running coupling constant $\alpha_s(\sim x P_z)$ and the 
perturbation series becomes a more physical expansion. 
But at small $x$, $\alpha_s(\sim x P_z)$ becomes very large and the perturbative matching breaks down. The physics is that if the quark/gluon longitudinal momentum $p_z$ is too small or comparable
to the intrinsic transverse momentum, it is no longer collinear with the hadron momentum so the factorization to infinite-momentum-frame parton physics breaks down. Thus, we expect a large discrepancy between RGR and fixed-order matching at small $x$. 
Moreover, we can compare resummations up to next-to-leading order (NLO) or next-to-next-to-leading order (NNLO) to check the convergence of perturbative expansions.

The above resummation has a correspondence in the coordinate space expansion~\cite{Braun:2007wv}. When
the quasi-PDF correlator is expanded in terms of moments of PDFs at short distances~\cite{Radyushkin:2017cyf,Ma:2017pxb,Izubuchi:2018srq}, the natural scale is the correlator distance $z$. This physical scale is recovered in the Wilson coefficients again through RGR. 
Operationally, $\sim 1/z$ shows up as an intrinsic scale for the running coupling constant $\alpha_s(\sim1/z)$. At large distances, $\alpha_s(\sim1/z)$ is too large, and the perturbation series breaks down. Only the correlation data at $z\ll 1/\Lambda_{\rm QCD}$ can be used
to extract the PDF moments. Therefore, we expect a very different conclusion 
about the moments fitting from those done with fixed-order perturbation theory~\cite{Radyushkin:2018cvn,Gao:2020ito}. 

Through applying RGR to the LaMET matching, our goal is to improve the accuracy of the fixed-order matching at intermediate $x$, and at the same time, to exhibit the limitation of perturbation theory at small $x$. 
We demonstrate the approach by using the lattice QCD matrix elements for pion valence PDF at lattice spacing $a$= 0.04\,fm and $a$= 0.06\,fm and momentum $P_{z}$= 0 $\sim$ 2.4 GeV calculated by BNL/ANL collaboration~\cite{Gao:2020ito,Gao:2021hxl,Gao:2021dbh}. Through
a similar resummation in coordinate space,
we also exhibit the large perturbative errors of Wilson coefficients at large distances, 
causing a large uncertainty in fitting $\langle x^4 \rangle$ in the OPE approach. 

The rest of this paper is organized as follows. In Sec.~\ref{sec:theory}, we present the formalism of renormalization group resummation in LaMET matching. In Sec.~\ref{sec:RGI}, we apply the RGR matching to pion valence PDF and compare it with the
fixed-order result. In Sec.~\ref{sec:wc&mom}, we study RGR effects in fitting the moments
of PDFs in the OPE approach using the same data. We summarize our results in Sec.~\ref{sec:summary}.

\section{Resumming quark's longitudinal momentum logarithms}\label{sec:theory}

In this section, we present the formalism for resumming quark's longitudinal momentum logarithms 
through the renormalization group in LaMET matching.

We start by renormalizing the lattice QCD matrix element $\tilde{h}^{\rm lat}(z, a, P_{z})=\langle P_z| \bar{\psi}(z) \gamma^{t} U(z,0) \psi (0) |P_z \rangle$ in the hybrid scheme~\cite{Ji:2020brr}:
\bea\label{eq:RenonpertHybrid1}
\tilde{h}(z, P_{z}) = \frac{\tilde{h}^{\rm lat}(z, a, P_{z})}{\tilde{h}^{\rm lat}(z, a, 0)}\theta(z_{s}-|z|) \nonumber\\
+ \frac{\tilde{h}^{\rm lat}(z, a, P_{z})}{Z^{R}(z,a,\mu) \tilde{h}^{\overline{\rm MS}}(z_{s}, \mu, 0)}\theta(|z|-z_{s}), 
\eea
where $z_{s}$ is the hybrid scheme parameter which satisfies $a \ll z_s \ll 1/\Lambda_{\rm QCD}$ to guarantee perturbation theory works when $z<z_s$.  $Z^{R}(z,a,\mu)$ is the phenomenological renormalization factor to eliminate linear and logarithmic divergences based on the self-renormalization method~\cite{LatticePartonCollaborationLPC:2021xdx}. It also contains the mass renormalization counterterm and scheme conversion factor to convert the lattice QCD matrix element to the $\overline{\rm MS}$ scheme. Therefore, we should have the following relation at short distances:
$$\frac{\tilde{h}^{\rm lat}(z, a, 0)}{Z^{R}(z,a,\mu)}=\tilde{h}^{\overline{\rm MS}}(z, \mu, 0),$$
where $z \ll 1/\Lambda_{\rm QCD}$. $\tilde{h}^{\overline{\rm MS}}(z, \mu, 0)$ is the zero momentum matrix element in the $\overline{\rm MS}$ scheme, which is perturbatively calculated up to NNLO in the literature~\cite{Li:2020xml}, see Eq.~\eqref{eq:MSbarpert}. We introduce $\tilde{h}^{\overline{\rm MS}}(z_{s}, \mu, 0)$ in the denominator at large distances in Eq.~\eqref{eq:RenonpertHybrid1} to guarantee that $\tilde{h}(z, P_z)$ is continuous,
although the derivative needs not be. An important feature of the hybrid renormalization is that at short distances, the ratio of
the lattice matrix elements has the proper normalization at $z=0$ with some of the discretization effects being canceled. While at large distances, no undesired non-perturbative effects are introduced. 

Fourier transforming the hybrid renormalized matrix element:
\bea\label{eq:FT}
&&\tilde{f}(x, P_{z}) = P_z \int_{-\infty}^{+\infty} \frac{d z}{2 \pi} e^{i x P_z z} \tilde{h}(z, P_{z}),
\eea
we obtain the quasi-PDF $\tilde{f}(x, P_{z})$ in the hybrid scheme, 
whose factorization in momentum space is:
\bea\label{eq:matching}
\tilde{f}(x, P_{z}) 
&&= \int_{-1}^{1} \frac{dy}{\left| y \right|} C\left(\frac{x}{y},\frac{\mu}{|x|P_{z}}\right) f(y, \mu) \nonumber\\
&&+ {\cal O}\left[\frac{\Lambda_{\rm QCD}^2}{x^2 P_{z}^2}, \frac{\Lambda_{\rm QCD}^2}{(1-x)^2 P_{z}^2}\right],
\eea
where $f(y, \mu)$ is the light-cone PDF and $C\left(\frac{x}{y},\frac{\mu}{|x|P_{z}}\right)$ is the perturbative matching kernel, which has been calculated up to NNLO in the $\overline{\rm MS}$ scheme for the isovector combination~\cite{Li:2020xml,Chen:2020ody}. Based on their calculations, we present the NLO and NNLO hybrid scheme matching kernels in Eqs.~\eqref{eq:hybridkernelNLO} and \eqref{eq:hybridkernelNNLO} in Appendix~\ref{app:hybridkernel}. Note that we write one of the arguments as $\frac{\mu}{|x|P_{z}}$ instead of $\frac{\mu}{|y|P_{z}}$ because we should have the momentum of the quasi parton $|x| P_z$ in the matching kernel~\cite{Izubuchi:2018srq}. 

In the matching kernel Eq.~\eqref{eq:ratiokernelNLO}, there is a logarithmic term $\sim \ln \left(\frac{\mu^2}{4 x^2 P_z^2} \right)$. We group $2x$ with $P_z$ because it appears that $2 x P_{z}$ is the ``intrinsic'' physical scale of the quasi-PDF. Apart from the natural combination appearing in the one-loop result, we can argue the case by an analogy with deep inelastic scattering (DIS), in which a virtual photon of momentum $q^\mu$ scatters on a nucleon target of momentum $P^\mu$, producing final states of high virtuality. The standard QCD factorization formula for one of the unpolarized DIS structure functions $F_2$ is as follows~\cite{Collins:2011zzd}:
\bea\label{eq:DISfact}
F_2(Q,x_B)=&&\int_{x_B -}^{1+} \mathrm{d} x \hat{C}_{2}\left(\frac{Q}{\mu}, \frac{x_B}{x}\right) f(x,\mu) \nonumber\\
&&+ {\cal O}\left[\frac{\Lambda_{\rm QCD}^2}{Q^2}\right],
\eea
where the summation over quark flavor and gluon has been omitted for simplicity, $Q^2=-q^2$ is the virtuality of the photon and $x_B = \frac{Q^2}{2 P \cdot q}$ is the Bjorken variable. $f(x,\mu)$ is the light-cone PDF and $\hat{C}_{2}\left(Q / \mu, x_B / x\right)$ is the coefficient function that matches the light cone PDF to the structure function, which was calculated up to NNLO~\cite{Zijlstra:1992qd,Moch:1999eb} and ${\rm N^{3}LO}$~\cite{Vermaseren:2005qc} in literature. We present the NLO quark result here~\cite{Collins:2011zzd}: 
\bea
&&\hat{C}_{2}(Q/\mu,z) = \delta(1-z) \nonumber\\
&&+ \frac{\alpha_s C_F}{2 \pi} z \left[\frac{1+z^2}{1-z}\left(-\ln \left[\frac{\mu^2}{Q^2}\right]+\ln \left[\frac{1-z}{z}\right]\right) \right. \nonumber\\
&&\left.-\frac{3}{2}\frac{1}{1-z}+3+2 z\right]_{+}.
\eea
There are logarithmic terms $\sim \ln(\frac{\mu^2}{Q^2})$ in the above formula and the higher-order coefficient functions. The scale $\mu$ dependence of these log terms in $\hat{C}_{2}$ cancels the scale $\mu$ dependence of the light cone PDF and the QCD running coupling so that the structure function $F_{2}$ is scale $\mu$ independent. For example, the leading log term $\frac{\alpha_s C_F}{2 \pi} z \frac{1+z^2}{1-z}\left(-\ln \left[\frac{\mu^2}{Q^2}\right]\right)$ in the NLO result is correspondent to the leading order Dokshitzer–Gribov–Lipatov–Altarelli–Parisi (DGLAP) evolution. These log terms are large if $\mu$ differs from $Q$ by orders of magnitude. To resolve this issue, it is natural to choose $\mu=Q$ in the quark coefficient function and extract PDFs at certain scales through DGLAP evolution.

\begin{figure}[tbp]
    \centering
    \includegraphics[width=.35\textwidth]{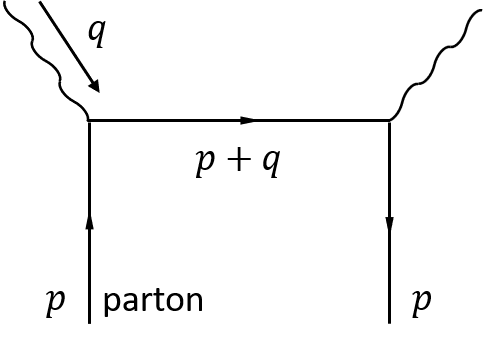}
    \caption{Kinematics of deep inelastic scattering in the parton model.}
    \label{fig:DISkinematics}
\end{figure} 

The factorization formula Eq.~\eqref{eq:DISfact} is Lorentz invariant and we consider a specific reference (Bjorken or Breit) frame for the purposes of understanding the physical scale. In this frame, the energy of virtual photon is zero and the momentum distribution
of the partons is measured (see Fig.~\ref{fig:DISkinematics}):
\bea
&&q^\mu=(0,0,0,-2 x P_z) \nonumber\\
&&p^\mu=(x P_z, 0, 0, x P_z) \nonumber\\
&&(p+q)^\mu =(x P_z, 0, 0, -x P_z).
\eea
Here we have rewritten the photon momentum in terms of the momentum of the struck parton. 
It is now clear that the virtuality of the photon is $Q^2=-q^2=(2 x P_z)^2$. Although the above kinematics is for tree level scattering, $Q^2=(2 x_B P_z)^2$ can be seen from the definition of the Bjorken variable $x_B = \frac{Q^2}{2 P \cdot q}$ in this frame where $P^{\mu}=(P_z,0,0,P_z)$ and $q^{\mu}=(0,0,0,-2 x_B P_z)$.

In Bjorken frame, it is natural to calculate the structure function in terms of the quasi-PDF, rather than the light-cone PDF. One can then recover the above standard DIS QCD factorization by matching quasi-PDF further to light-cone one as in Eq.~\eqref{eq:matching}. This new approach of using quasi-PDF may help us understand the origin of its physical scale, 
\bea\label{eq:DISfactBj}
F_2(Q,x_B) = \hat{\tilde{C}}_{2}\left(\frac{Q}{\mu},\frac{x_B}{x}\right) \otimes \tilde{f}(x, P_{z}) 
\eea
where $\hat{\tilde{C}}_{2}\left(\frac{Q}{\mu},\frac{x_B}{x}\right)$ is the quasi DIS coefficient function that matches the quasi-PDF to the structure function. Note that the fraction $\frac{Q}{2 x P_z}$ is just $\frac{x_B}{x}$. $\hat{\tilde{C}}_{2}$ doesn't contain the DGLAP logs if the quasi-PDF is scale $\mu$ independent as in hybrid scheme~\cite{Ji:2020brr}. For example, the NLO result for $\hat{\tilde{C}}_{2}$ during the region $0 < x_B/x < 1$ is presented here
\bea
&&\hat{\tilde{C}}_{2}\left(\frac{Q}{\mu},z\right) = \delta(1-z) + \frac{\alpha_s C_F}{2 \pi} z \left[ -\frac{z^2}{1-z} \right]_{+} + ...
\eea
where $...$ denotes the counterterm corresponding to the renormalization scheme of the quasi-PDF. If we renormalize the quasi-PDF in the hybrid scheme~\cite{Ji:2020brr}, the counterterm doesn't contain any new logs, and thus the NLO result doesn't contain $\sim \ln(\frac{\mu^2}{Q^2})$. There is $\sim \beta_0 \ln(\frac{\mu^2}{Q^2})$ at NNLO, which is the subleading log to cancel the scale $\mu$ dependence of the running coupling $\alpha_s(\mu)$.

Through comparing Eqs.~\eqref{eq:matching},~\eqref{eq:DISfact} and~\eqref{eq:DISfactBj}, we obtain
\bea
\hat{C}_{2}\left(\frac{2 x_B P_z}{\mu}, \frac{x_B}{y} \right)=\hat{\tilde{C}}_{2}\left(\frac{2 x_B P_z}{\mu},\frac{x_B}{x}\right) \otimes C\left(\frac{x}{y},\frac{\mu}{|x|P_{z}}\right).\nonumber\\
\eea
The DGLAP logs on both sides of this equation should match. On the left hand side, the DGLAP logs behave as $\sim \ln \left[\frac{\mu^2}{(2 x_B P_z)^2}\right]$. On the right hand side, since $\hat{\tilde{C}}_{2}$ doesn't contain the DGLAP logs, the DGLAP logs only come from the LaMET matching kernel $C$, which should behave as $\sim \ln \left[\frac{\mu^2}{(2 x P_z)^2}\right]$. Convoluting $\sim \ln \left[\frac{\mu^2}{(2 x P_z)^2}\right]$ in $C$ with $\hat{\tilde{C}}_{2}$, we obtain the DGLAP logs $\sim \ln \left[\frac{\mu^2}{(2 x_B P_z)^2}\right]$ in $\hat{C}_{2}$. For example, convoluting the leading logs $\sim \ln \left[\frac{\mu^2}{(2 x P_z)^2}\right]$ in $C$ with the $\delta(1-x_B/x)$ in LO $\hat{\tilde{C}}_{2}$, we obtain the leading logs $\sim \ln \left[\frac{\mu^2}{(2 x_B P_z)^2}\right]$ in $\hat{C}_{2}$.

Since the DGLAP logs in the LaMET matching kernel are $\sim \ln \left[\frac{\mu^2}{(2 x P_z)^2}\right]$, the intrinsic physical scale of the quasi-PDF is $2 x P_z$. Therefore,
we introduce a notation $Q_{\rm eff}=2xP_zc'$ for the quasi-PDF matching, with $c'\sim 1$. For a fixed $P_z$, the scale becomes small at small $x$, and the resulting logarithms in the matching kernel become large.

We need to resum this type of large logarithms $\sim \ln \left(\frac{\mu^2}{4 x^2 P_z^2} \right)$ to all orders in perturbation theory 
to recover the small-$x$ behavior
in the LaMET calculations. The main result is that the natural scale for the perturbative expansion is $2xP_z$, which appears in the coupling $\alpha_s(2 x P_z)$. This indicates that perturbative matching is not valid for small $x$ because the meaning of parton is lost for a quark with small longitudinal momentum
at the order of $\Lambda_{\rm QCD}$. In fact, the same conclusion arises from the
higher-twist power counting~\cite{Ji:2020ect}.
In the factorization formula Eq.~\eqref{eq:matching}, part of the higher twist effects behave like ${\cal O}\left[\frac{\Lambda_{\rm QCD}^2}{x^2 P_{z}^2}\right]$, which is also large for small $x$. 

The small-momentum large logarithms can be resummed using
the standard renormalization group method. 
One can invert the matching kernel to obtain the light cone PDF:
\bea\label{eq:inversematching}
f(x,\mu) =&& C^{-1}\left(\frac{x}{y}, \frac{\mu}{|x| P_{z}}\right) \otimes \tilde{f}(y,P_z) \nonumber\\
&&+ {\cal O}\left[\frac{\Lambda_{\rm QCD}^2}{x^2 P_{z}^2}, \frac{\Lambda_{\rm QCD}^2}{(1-x)^2 P_{z}^2}\right],
\eea
where the convolution is now performed with respect to $y$. The inverse matching kernel $C^{-1}\left(\frac{x}{y}, \frac{\mu}{|x| P_{z}}\right)$ is now expanded in terms of $\alpha_s(\mu)$. After we invert the matching kernel, $x$ in the logarithmic term $\ln \left(\frac{\mu^2}{4 x^2 P_z^2} \right)$ and higher-twist effects becomes the momentum fraction of a light-cone parton, which is shown in Appendix~\ref{app:inversematching}. Taking the derivative with respect to scale $\mu$ on both sides of Eq.~\eqref{eq:inversematching}, and noticing that $\tilde{f}(y,P_z)$ is scale independent,
we obtain the renormalization group equation for $C^{-1}\left(\frac{x}{y}, \frac{\mu}{|x| P_{z}}\right)$, which is the same as that for $f(x,\mu)$:
\bea\label{eq:matchingRGI}
&&\mu \frac{d C^{-1}\left(\frac{x}{y},\frac{\mu}{|x| P_z}\right)}{d\mu} \nonumber\\ 
&&= \int_{x}^{1} \frac{dw}{w} P[w,\alpha_{s}(\mu)] C^{-1}\left(\frac{x/w}{y},\frac{\mu}{|x/w| P_z}\right), 
\eea
where $P[w,\alpha_{s}(\mu)]$ is the DGLAP evolution kernel of the light-cone PDF:
\bea\label{eq:DGLAP}
\mu \frac{d f(x,\mu)}{d \mu} = \int_{x}^{1} \frac{d w}{w} P[w,\alpha_{s}(\mu)]f\left(\frac{x}{w},\mu\right),
\eea
which has been calculated up to three loops in literature~\cite{Moch:2004pa}. 

To get the RGR matching kernel $C_{\rm RGR}\left(\frac{x}{y},\frac{\mu}{|x|P_{z}}\right)$, we can solve Eq.~\eqref{eq:matchingRGI} and evolve to scale $\mu$, where the initial condition at scale $Q_{\rm eff}=2 x P_{z} c'$ is the fixed order matching kernel $C$: 
$C^{-1}_{\rm RGR}\left(\frac{x}{y},\frac{Q_{\rm eff}}{|x|P_{z}}\right)\approx C^{-1}\left(\frac{x}{y},\frac{Q_{\rm eff}}{|x|P_{z}}\right).$
We can vary $c'$ around 1 to test the stability of perturbative matching. The resummation scale is an order-of-magnitude estimate based on the physical scale and can vary without having meaningful physics impact, hence the parameter $c'$. Different choices of
resummation scale ($c'$) correspond to higher order effects in $\alpha_s$, which disappear when the perturbation series are computed to the infinite order.
If the effect of varying $c'$ is small, the higher order effect is small and the perturbation series has a good convergence. 
So the variation of $c'$ can be used to test the stability (convergence) of perturbation series. However, this is necessary but not sufficient as we have not actually calculated the higher-order terms. 
The variation limits of $c'$ are conventional. When people study the resummation effect near the Z boson mass $m_Z$, people usually vary $c'=0.5\sim2$, which leads to the variation of $\alpha_s(c'm_Z)$ for about 10\%. When we study the perturbation series near $\sim 2~$GeV, to have the 10\% variation of $\alpha_s$, we choose $c'=0.8\sim1.2$.

In practical calculations, it is complicated to solve Eq.~\eqref{eq:matchingRGI} directly because it contains plus functions in both the DGLAP kernel and the matching kernel. Here we 
present an equivalent approach. First of all, we can
calculate PDF at its own intrinsic scale $Q_{\rm eff}=2 x P_z c'$:
\bea
&&f(x,Q_{\rm eff}) = C^{-1}\left(\frac{x}{y}, \frac{Q_{\rm eff}}{|x| P_{z}}\right) \otimes \tilde{f}(y,P_z),
\eea
where $C^{-1}$ is the inverse of the fixed order matching kernel (the scale of $\alpha_s$ in $C^{-1}\left(\frac{x}{y}, \frac{Q_{\rm eff}}{|x| P_{z}}\right)$ is $Q_{\rm eff}$). Then we just need to DGLAP evolve $f(x,Q_{\rm eff})$ to scale $\mu$. Note that this is an unusual DGLAP evolution since the scales in the initial PDF $f(x,2 x P_z c')$ are different for different $x$. A method for such a DGLAP evolution is provided in Appendix~\ref{app:exactscalemethod}.

\section{Resummation Effect in Pion PDF Calculation}\label{sec:RGI}

As an example, we study the effect of resumming the small-momentum large
logarithms in the pion valence PDF calculation.  
The resummation improves the lattice QCD prediction
in the $2 x P_z$ range 0.7 GeV to 1.5 GeV, but generates a 
very large uncertainty in the smaller-$x$ region. 

\subsection{Lattice Data and Renormalization}\label{sec:latticedata}
Lattice QCD matrix elements for the valence parton distribution of a pion at lattice spacing $a$= 0.04\,fm and $a$= 0.06\,fm and momentum $P_{z}$= 0 $\sim$ 2.4 GeV are calculated by the BNL/ANL collaboration. These data have been analyzed in recent BNL/ANL papers~\cite{Gao:2020ito,Gao:2021hxl,Gao:2021dbh}. A detailed explanation of their data can be found in Ref.~\cite{Gao:2020ito}. Here is the definition of their lattice matrix elements:
\bea\label{eq:latM}
&&\tilde{h}^{\rm lat}(z, a, P_{z}) = \\ 
&&\langle \pi^{+}(P_z)| \bar{u}(z) \gamma^{t} U(z,0) u(0) - \bar{d}(z) \gamma^{t} U(z,0) d(0) |\pi^{+}(P_z) \rangle, \nonumber
\eea
where the values of momenta $P_{z}$ at two lattice spacings are listed in Table~\ref{tab:latticedata}. Notice that the isovector distribution $u(x)-d(x)$ is the same as the valence distribution $u(x)-\bar{u}(x)$ for pion because of the isospin symmetry. 

\begin{table}[htbp]
\centering
    \begin{tabular}{|c|c|c|}
    \hline \hline $n_{z}$ & 
    \multicolumn{2}{|c|}{$P_{z}(\mathrm{GeV})$} \\
    \cline { 2 - 3 } & $a=0.06 \mathrm{fm}$ & $a=0.04 \mathrm{fm}$ \\
    \hline 0 & 0 & 0 \\
    1 & $0.43$ & $0.48$ \\
    2 & $0.86$ & $0.97$ \\
    3 & $1.29$ & $1.45$ \\
    4 & $1.72$ & $1.93$ \\
    5 & $2.15$ & $2.42$ \\
    \hline \hline
    \end{tabular}
    \caption{The values of momenta $P_{z}=\frac{2 \pi}{L a} n_{z}$ at two lattice spacings~\cite{Gao:2020ito}.}
    \label{tab:latticedata}
\end{table}

The matrix elements are calculated on two different ensembles $L_t \times L^3$: One ensemble has lattice spacing $a=0.06$~fm and lattice volume $48 \times 64^3$, and the other one has lattice spacing $a=0.04$~fm and lattice volume $64 \times 64^3$. These ensembles were generated by HotQCD collaboration~\cite{HotQCD:2014kol} with 2+1 flavor Highly Improved Staggered Quarks (HISQ) in the sea~\cite{Follana:2006rc}. One step of HYP smearing is applied on the gauge links, which are used in both quark propagators and Wilson links in the quasi-PDF operator. Wilson clover quark action is used during the calculation. They tune the quark mass so that the valence pion mass is 300 MeV. 

When extracting the ground state pion matrix elements Eq.~\eqref{eq:latM}, they perform different fitting strategies, including fitting the ratio and fitting the summed ratio. They get consistent results among different fitting strategies. 

We parametrize the renormalization factor as
\bea\label{ZRmodel}
Z^{R}(z,a,\mu) = \exp{[-\delta m(a) z - m_{0}^{\rm eff}z + b(a,\mu)]},
\eea
where $\delta m(a)$ is the linear divergence coefficient, which is extracted from static quark-antiquark potential in~\cite{Gao:2021dbh}: $a \delta m(a=0.06~{\rm fm})=0.1586(8)$ and $a \delta m(a=0.04~{\rm fm})=0.1508(12)$. We need to consider an effective renormalon ambiguity $m_0^{\rm eff}$ because we cannot make sure that we regularize the lattice perturbation series and $\overline{\rm MS}$ perturbation series in the same way, and we also include part of the non-perturbative contribution in $m_0^{\rm eff}$, which is to convert the lattice matrix element to $\overline{\rm MS}$ scheme up to the linear $z$ accuracy. We expect that $m_0^{\rm eff}$ is independent of the fit range $z$. The term $b(a,\mu)$ is for logarithmic divergence, scheme conversion factor and scale dependence. 

\begin{figure}[htbp]
    \centering
    \includegraphics[width=8.5cm]{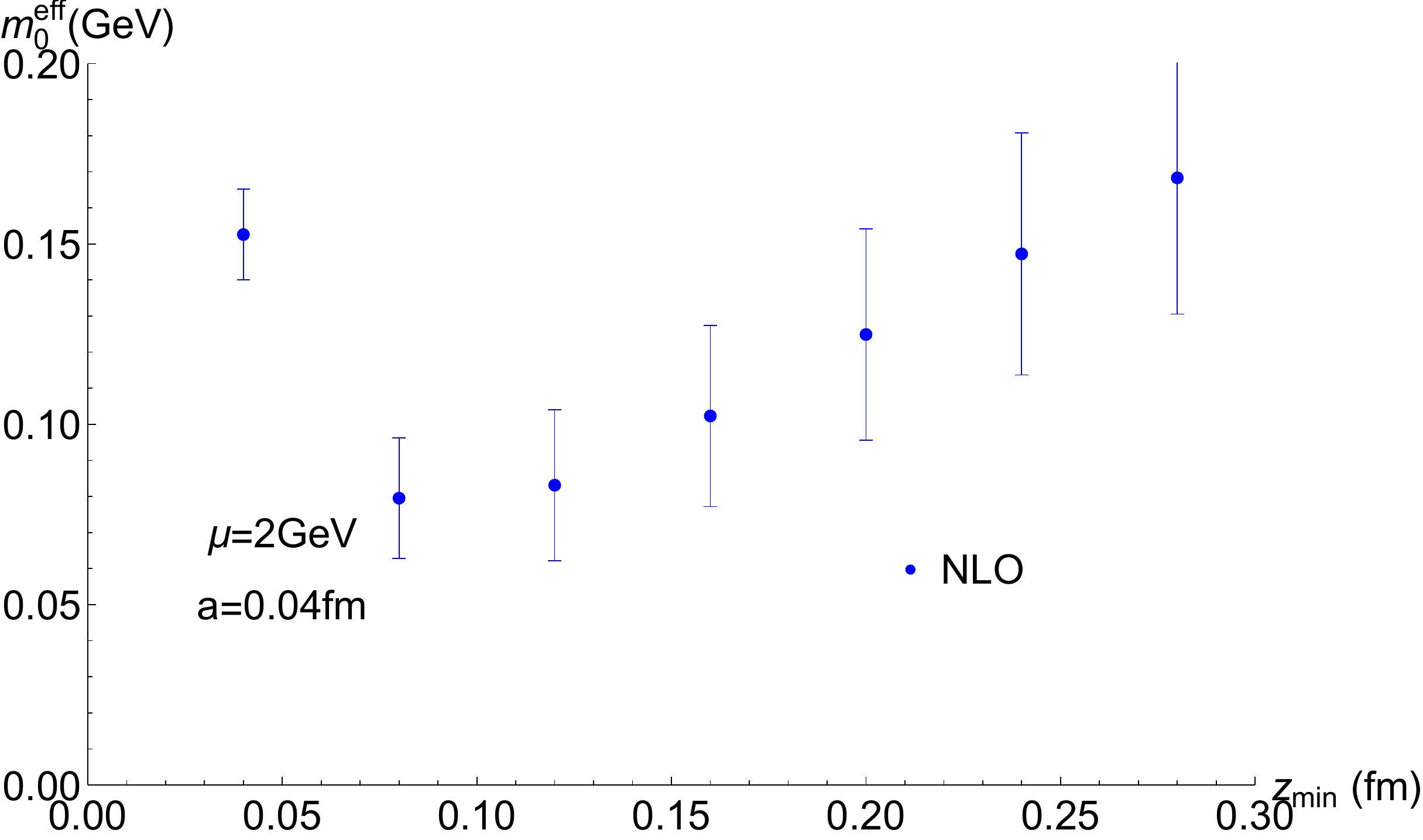}
    \caption{Fitted $m_0^{\rm eff}$ with respect to $z_{\rm min}$, based on NLO $\tilde{h}^{\overline{\rm MS}}(z, \mu, 0)$.}
    \label{fig:m0effNLO}
\end{figure}
We fit $m_0^{\rm eff}$ and $b(a,\mu)$ through requiring that the renormalized zero momentum matrix element is equal to the perturbation series $\tilde{h}^{\overline{\rm MS}}(z, \mu, 0)$ (Eq.~\eqref{eq:MSbarpert}) at short distances:
$$
\tilde{h}^{\overline{\rm MS}}(z, \mu, 0) = \frac{\tilde{h}^{\rm lat}(z, a, 0)}{Z^{R}(z,a,\mu)},
$$
where $z \ll 1/\Lambda_{\rm QCD}$. For $a=0.04~{\rm fm}$, we fit in the range $z \in [z_{\rm min}, z_{\rm min}+0.08~{\rm fm}]$ with NLO $\tilde{h}^{\overline{\rm MS}}(z, \mu, 0)$ and the fitted $m_0^{\rm eff}$ with respect to $z_{\rm min}$ is shown in Fig.~\ref{fig:m0effNLO}. In this figure, the fitted $m_0^{\rm eff}$ depends on $z_{\rm min}$.
However, for our demonstrative purpose, we just use the $m_0^{\rm eff}$ from the fit range $z \in [0.08,0.16]$~fm, which is $0.080(17)$ GeV. And we will use the same $Z^{R}(z,a,\mu)$ extracted based on NLO $\tilde{h}^{\overline{\rm MS}}(z, \mu, 0)$ for all the cases in this section.

\begin{figure}[tbp]
    \centering
    \includegraphics[width=8.5cm]{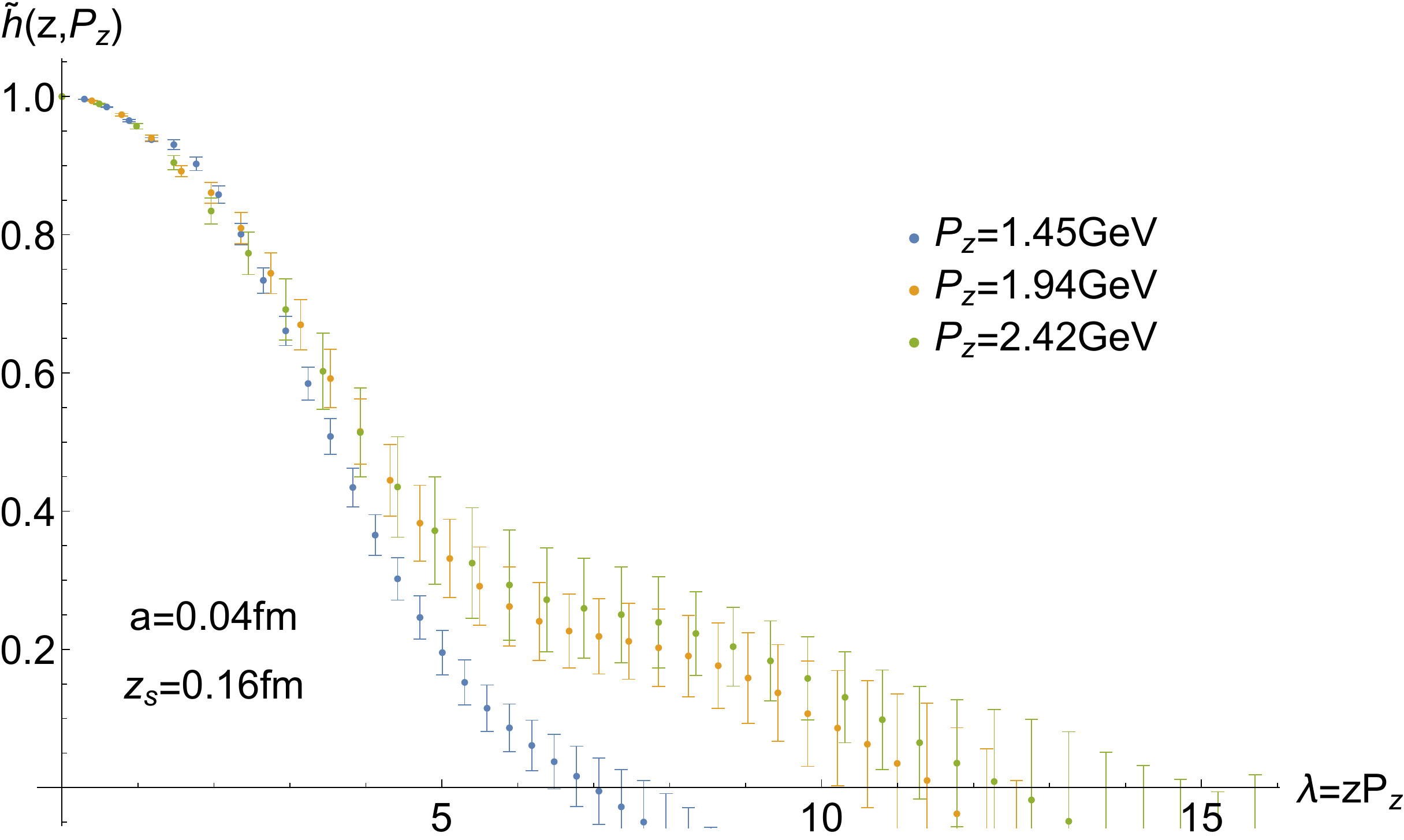}
    \caption{Renormalized lattice matrix elements in the hybrid scheme (colorful points). $a=0.04$~fm and $z_s=0.16$~fm. }
    \label{fig:hhybridnonpert}
\end{figure}
Following Eq.~\eqref{eq:RenonpertHybrid1}, we renormalize the lattice QCD matrix element in the hybrid scheme, and the result is shown in Fig.~\ref{fig:hhybridnonpert}. This is not too different from the renormalized data in Ref.~\cite{Gao:2021dbh} because we use the same linear divergence factor $\delta m$, which has the largest influence during the renormalization process. The differences are just higher order power corrections such as $O(z^2_s)$ which are small (less than 10\%) if $z_s$ is a perturbative short distance.

We do extrapolation on the hybrid renormalized matrix element with the following form suggested by Ref.~\cite{Ji:2020brr}: 
$$
\tilde{h}(\lambda) = \left[\frac{c_1}{\left| \lambda \right| ^{d_1}} \cos \left(\frac{\pi  d_1}{2}\right)+\frac{c_2}{\left| \lambda \right| ^{d_2}} \cos \left(\left| \lambda \right| -\frac{\pi  d_2}{2}\right)\right] e^{-\frac{\left| \lambda \right| }{\lambda _0}}
$$   
where $\lambda=zP_z$. Then we do a Fourier transformation on the hybrid renormalized matrix element based on Eq.~\eqref{eq:FT}, where we use the extrapolation after $\lambda=9$. The above form comes from the asymptotic behavior of PDF near the end point region $x^{a}(1-x)^{b}$. We Fourier transform it and obtain the coordinate space distribution. At large $\lambda$, we obtain the form $\left[\frac{c_1}{\left| \lambda \right| ^{d_1}} \cos \left(\frac{\pi  d_1}{2}\right)+\frac{c_2}{\left| \lambda \right| ^{d_2}} \cos \left(\left| \lambda \right| -\frac{\pi  d_2}{2}\right)\right]$. Since it is a form for quasi-PDF with finite $P_z$, there may be a mass gap so we can introduce an exponential decay $e^{-\frac{\left| \lambda \right| }{\lambda _0}}$ in the form. 

\subsection{LaMET Result with RG-Resummation to NLO and NNLO}

In this section, we perform RGR matching based on the method in Appendix~\ref{app:exactscalemethod}.

\begin{figure}[htbp]
\centering
    \includegraphics[width=8.5cm]{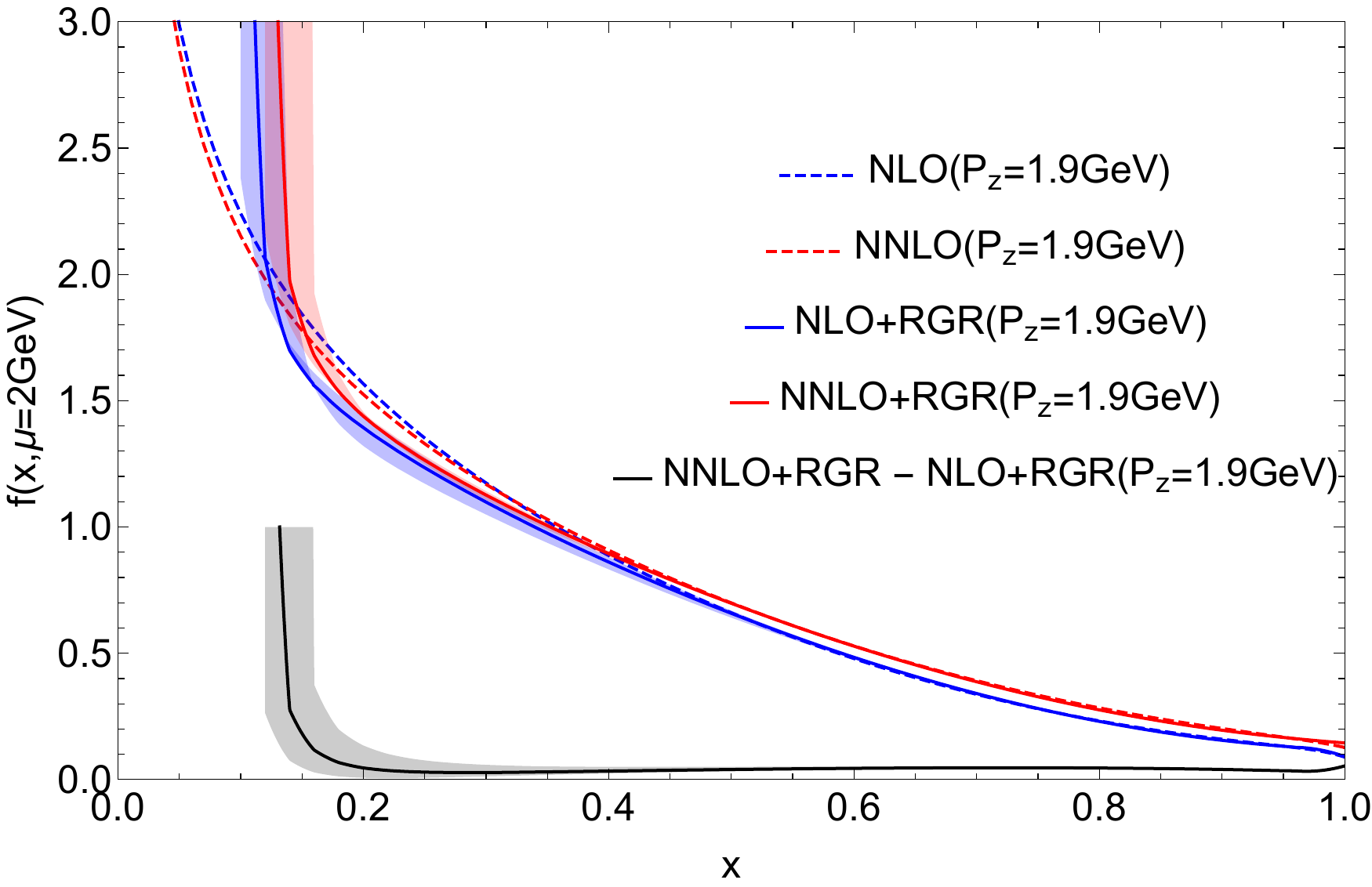}
    \caption{Pion valence PDFs calculated with fixed-order matching kernel (dashed) and RGR matching (solid, blue and red). The black curve shows the absolute difference between NNLO+RGR and NLO+RGR PDFs. The lattice data is at $P_z=1.9$~GeV and $a=0.04$~fm. The bands show the uncertainty from varying $c'=0.8\sim1.2$.}
    \label{fig:PDFdifferentkernels}
\end{figure}

The RGR matched PDFs for $P_z=1.9$~GeV are shown in Fig.~\ref{fig:PDFdifferentkernels}, in comparison with fixed-order matched PDFs. In the large $2 x P_z$ region (e.g. $2 x P_z >$ 2.7~GeV), there are no qualitative differences between RGR and fixed order matched PDFs in this case. The reason is that the physical scale $2 x P_z$ for $P_z=1.9$~GeV at large $x$ region is close to the PDF scale $\mu=2$~GeV we choose in the matching, where the resummed logs are not very large. However, if we have much larger momenta (e.g. $P_z = 10~{\rm GeV}$), we would expect to observe the resummation effects in the large $x$ region. In the moderate $2 x P_z$ region (e.g. $1.5~{\rm GeV}<2 x P_z<2.7~{\rm GeV}$), RGR matched and fixed-order matched PDFs are consistent with each other, which indicates that fixed-order perturbative matching works well. In the intermediate region (e.g. $0.7~{\rm GeV}<2 x P_z<1.5~{\rm GeV}$), resummation effects start to become important and improve the accuracy of the theoretical prediction. At small $2 x P_z$ (e.g. $2 x P_z<0.7~{\rm GeV}$), the effective coupling $\alpha_s(2 x P_z)$ becomes too large to be perturbative, and there are large discrepancies between RGR matched and fixed-order matched PDFs. There is also a large discrepancy between the NLO+RGR PDF and the NNLO+RGR PDF, indicating higher-order effects are not negligible. Therefore, the resummation of the large logarithms makes clear the breakdown of the perturbative matching at small $2 x P_z$ (e.g. $2 x P_z<0.7~{\rm GeV}$), where higher-twist effects are also unmanageable.

\begin{figure}[tbp]
\centering
    \includegraphics[width=8.5cm]{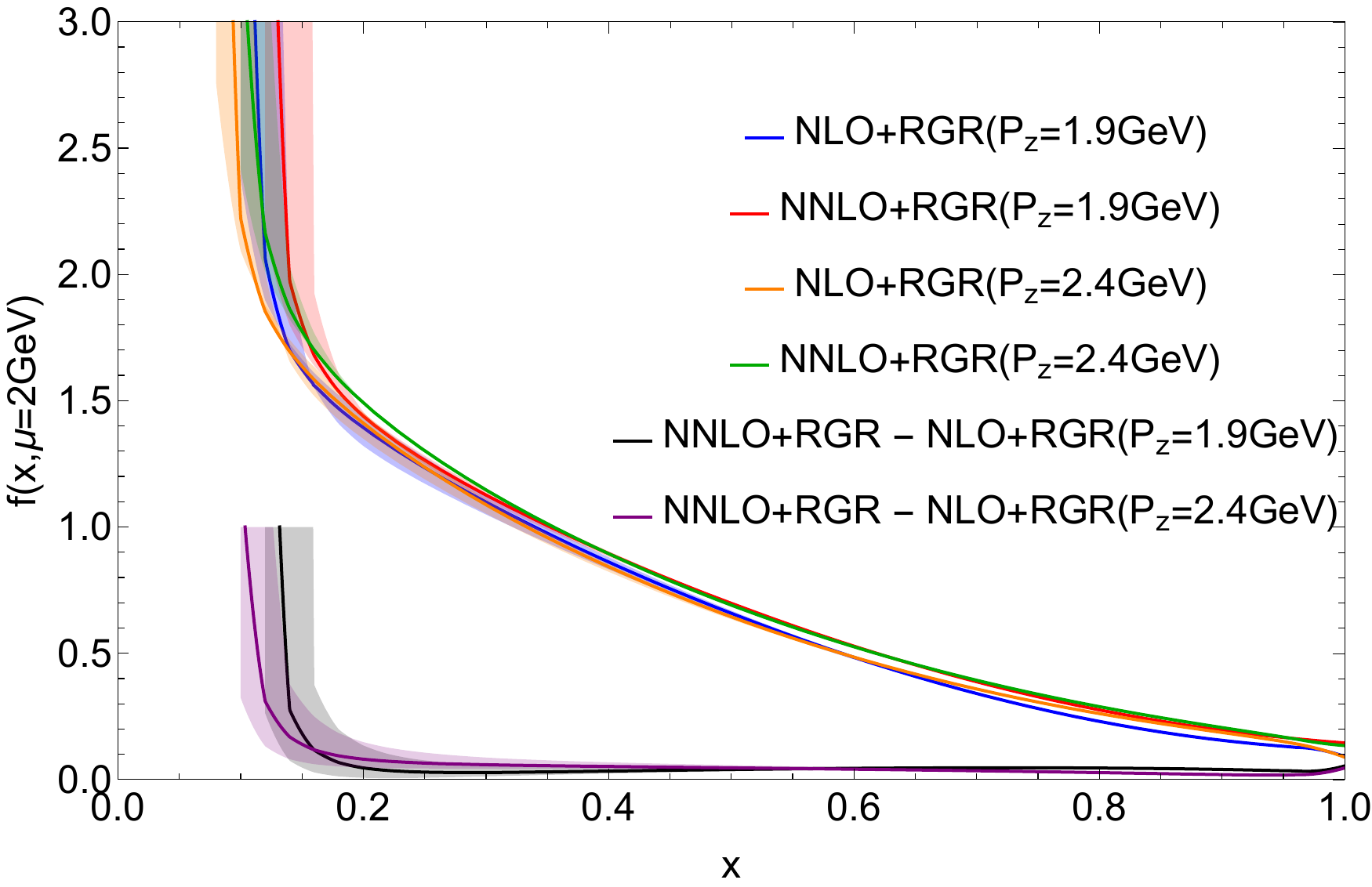}
    \caption{ Comparison between pion valence PDFs calculated with RGR matching at $P_z=1.9$~GeV (blue and red) and $2.4$~GeV (orange and green). The black and purple curves show the absolute difference between NNLO+RGR and NLO+RGR PDFs for $P_z=1.9$~GeV and $P_z=2.4$~GeV respectively. The bands show the uncertainty from varying $c'=0.8\sim1.2$. Better convergence is seen at larger momentum.}
    \label{fig:PDFdifferentmoms}
\end{figure}

The perturbative matching breaks down for $x < \frac{0.7~{\rm GeV}}{2 P_z}$ in the consideration of perturbative convergence. So with the availability of larger $P_z$ we will be able to reliably extract light-cone PDFs at smaller $x$. We further compare RGR matched PDFs for different $P_z$ in Fig.~\ref{fig:PDFdifferentmoms}. As expected, the discrepancy between the NNLO+RGR PDF and the NLO+RGR PDF in the small $x$ region (e.g. $x<0.16$) is smaller for $P_z=2.4$~GeV than that for $P_z=1.9$~GeV. Obviously, the convergence of perturbative matching at small $x$ can be improved systematically if we increase the hadron momentum.

\begin{figure}[tbp]
\centering
    \includegraphics[width=8.5cm]{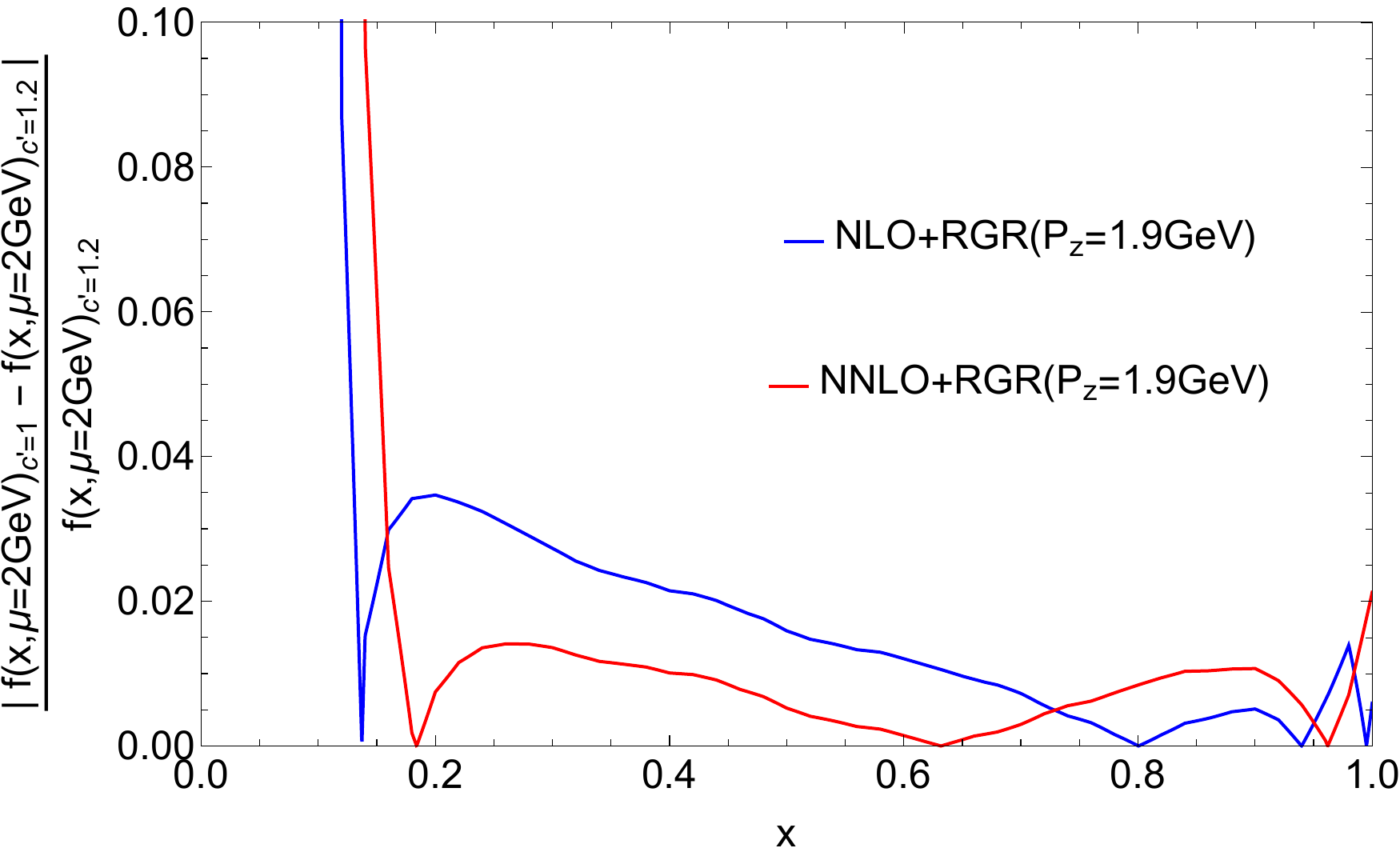}
    \caption{The relative difference between RGR matched PDFs at $c'=1$ and $c'=1.2$. The lattice data is at $P_z=1.9$~GeV and $a=0.04$~fm. Convergence is observed at moderate $x$ and divergence is observed at small $x$.}
    \label{fig:PDFsyserr}
\end{figure}

As shown in Fig.~\ref{fig:PDFsyserr}, the relative error caused by varying $c'$ for NNLO+RGR is smaller than that for NLO+RGR in the intermediate or moderate $x$ region (e.g. $0.6~{\rm GeV}<2 x P_z<2.8~{\rm GeV}$). Because the effect of varying $c'$ is a higher $\alpha_s$ order effect and the $\alpha_s(2 x P_z)$ is small in the intermediate- or moderate-$x$ region, we expect to get a smaller truncation error as the order of the perturbation series increases.
On the other hand, for $2 x P_z<0.6~{\rm GeV}$, the error in NNLO+RGR is larger than that in NLO+RGR. Since $\alpha_s(2 x P_z)$ is large at small $x$, a perturbative expansion becomes less convergent at this coupling and the NLO+RGR and NNLO+RGR results can differ considerably. 

Our main conclusions above will not change if we use a different $\Gamma$ matrix in the quasi-PDF operator. The intrinsic physical scale $2 x P_z$ follows directly from our analogy with DIS process, which is independent of $\Gamma$ matrix. And for the unpolarized PDF cases, the log term $\sim \ln \left(\frac{\mu^2}{4 x^2 P_z^2} \right)$ in the matching kernel is independent of $\Gamma$ matrix~\cite{Izubuchi:2018srq}.

\section{Resummed Coefficients in Operator Product Expansion and Moment Fit}\label{sec:wc&mom}

The connection between the Euclidean correlator in Eq.~\eqref{eq:quasi} and PDF has also been
explored through coordinate-space operator product expansion (OPE)~\cite{Radyushkin:2017cyf,Braun:2007wv,Izubuchi:2018srq}. 
With a finite-range coordinate-space correlation calculated from lattice 
QCD, it has been proposed to extract moments of PDFs through fitting 
the truncated OPE expression~\cite{Joo:2019bzr,Fan:2020nzz,Gao:2020ito}. To extract the moments reliably, one has to properly resum
large logarithms of type $\alpha^n(\mu)\ln^n(z\mu)$ in the Wilson coefficient~\cite{Gao:2021hxl}. Once
this is done, it is expected that OPE works only at perturbatively-small distances 
$z\ll 1/\Lambda_{\rm QCD}$ up to $0.2 \sim 0.3$~fm. This is conjugate to the
momentum-space expansion condition $2xP^z \gg \Lambda_{\rm QCD}$. We 
will show that with a $P^z\sim 2-3$ GeV, the pion correlator used in the previous 
section can determine the first moment $\langle x^2\rangle$ reliably, but $\langle x^4 \rangle$ marginally. 

The BNL lattice group has already done extensive analysis to extract
the pion PDF moments from lattice quasi-PDF correlators~\cite{Gao:2020ito,Gao:2021hxl,Gao:2022iex}.
In Ref.~\cite{Gao:2020ito}, the lattice correlator has been used to extract moments based on fixed-order Wilson coefficients, and they seem to get stable results for $\langle x^2 \rangle$, $\langle x^4 \rangle$ and $\langle x^6 \rangle$ when using lattice data 
up to $z$ = 0.6 fm or 0.8 fm. In Ref.~\cite{Gao:2021hxl}, they fit the data with the Wilson coefficients with NLO renormalization group resummation and threshold resummation, and they get good results for $\langle x^2 \rangle$ and $\langle x^4 \rangle$ when using data up to $z$ = 0.48 fm. In Ref.~\cite{Gao:2022iex}, they use data at physical pion mass and NNLO Wilson coefficients up to $z$ = 0.72 fm with fixed-order Wilson coefficients. They have studied whether RG-improved Wilson coefficients can describe lattice data in their Appendix B but they haven't estimated the uncertainties for moment fit from the systematic errors of Wilson coefficients at large distance.

Here, we repeat their analysis using RG resummed Wilson coefficients up to NNLO. We will illustrate the systematic errors of Wilson coefficients at large distance through varying
parameter $c$ in the resummation scale $c/z$. Our results show that the
perturbative Wilson coefficients are well-defined only up to about 0.3 fm, and as
a consequence, the resulting $\langle x^4 \rangle$ has a significant error. 

\subsection{Resummed Wilson Coefficients}

Following Refs.~\cite{Radyushkin:2017cyf,Orginos:2017kos,Radyushkin:2017lvu,Joo:2019jct,Joo:2019bzr,Fan:2020nzz,Gao:2020ito}, we consider a ratio of the lattice matrix element, which eliminates the linear divergence and logarithmic divergence:
\bea\label{eq:ratio}
\tilde{h}^{\rm ratio}(z, P_{z}, P_{z}^0) = \frac{\tilde{h}^{\rm lat}(z, a, P_{z})}{\tilde{h}^{\rm lat}(z, a, P_{z}^0)}.
\eea
We can relate $\tilde{h}^{\rm ratio}(z, P_{z}, P_{z}^0)$ with $\overline{\rm MS}$ scheme. Since the UV divergence is eliminated in the ratio and the intrinsic $z$ dependence should be the same between different schemes, we have
\bea\label{eq:ratioMSbarfixedorder}
\frac{\tilde{h}^{\rm lat}(z, a, P_{z})}{\tilde{h}^{\rm lat}(z, a, P_{z}^0)} = \frac{\tilde{h}^{\overline{\rm MS}}(z, \mu, P_{z})}{\tilde{h}^{\overline{\rm MS}}(z, \mu, P_{z}^0)},
\eea
where $\tilde{h}^{\overline{\rm MS}}(z, \mu, P_{z})$ is the renormalized lattice matrix element in $\overline{\rm MS}$ scheme.


According to OPE~\cite{Radyushkin:2017cyf,Braun:2007wv,Izubuchi:2018srq}, the coordinate space correlator can be related to the moments of PDF
at small $z\ll 1/\Lambda_{\rm QCD}$, 
\bea\label{eq:OPE}
\tilde{h}^{\overline{\rm MS}}(z, \mu, P_{z}) 
=&&\sum_{N=0}\frac{(-i\lambda)^N}{N!} C^{\overline{\rm MS}}_{N}(\alpha_{s},z^2 \mu^2) \, \langle x^N \rangle(\mu) \nonumber\\
&&+ {\cal O}(z^2 \Lambda_{\rm QCD}^2), 
\eea
where the moment $\langle x^N \rangle(\mu)=\int_{-1}^{1}dx \, x^N f(x,\mu)$.
Note that $C^{\overline{\rm MS}}_{0}(\alpha_{s},z^2 \mu^2)$ is equal to $\tilde{h}^{\overline{\rm MS}}(z, \mu, 0)$ in Eq.~\eqref{eq:MSbarpert} according to our definition. $C^{\overline{\rm MS}}_{N}(\alpha_{s},z^2 \mu^2)$ is calculated up to NNLO in the literature~\cite{Li:2020xml}. We present it in Appendix~\ref{app:Wilcoe}, see Eq.~\eqref{eq:WilcoeNLO} and Eq.~\eqref{eq:WilcoeNNLO}.

The renormalization scale $\mu$ is fixed, and usually chosen to be around 2 GeV. 
On the other hand, the correlation distance $z$ varies considerably. As such, 
the perturbative Wilson coefficient $C_N$ contains large logarithms of $\ln^n(z\mu)$ which must be resummed to get reliable results. Fortunately, this can be accomplished through 
solving the renormalization group equation for Wilson coefficient:
\bea\label{eq:RGEWC}
\mu \frac{d C^{\overline{\rm MS}}_{N}(\alpha_{s},z^2 \mu^2)}{d \mu} = \gamma^{\overline{\rm MS}}_{N}(\alpha_{s}) C^{\overline{\rm MS}}_{N}(\alpha_{s},z^2 \mu^2),
\eea
where $\gamma^{\overline{\rm MS}}_{N}$ is the anomalous dimension for the $N$-th Wilson coefficient, see Eq.~\eqref{eq:ADWilcoe}. 
The natural scale for the Wilson coefficient is the physical distance $z$, or 
in the momentum scale, $c/z$, with $c$ an order 1 number. Thus  we can sum over large
logarithms by integrating the above equation from $c/z$ to $\mu$, 
\bea\label{eq:WCRGI}
&&C^{\overline{\rm MS}{\rm RGR}}_{N}(\alpha_{s}(\mu),z^2 \mu^2) \nonumber\\
&&= C^{\overline{\rm MS}}_{N}(\alpha_{s}(c/z),c^2) \exp{\left[\int_{\alpha_{s}(c/z)}^{\alpha_{s}(\mu)} \frac{\gamma^{\overline{\rm MS}}_{N}(\alpha_{s})}{\beta^{\overline{\rm MS}}(\alpha_{s})} d \alpha_{s}\right]},
\eea
where $\beta^{\overline{\rm MS}}$ is the QCD beta function~\cite{Zyla:2020zbs}, and $c$ can be varied around $2e^{-\gamma_{E}}$ to test the stability of Wilson coefficient. 
Thus, the natural moment scale for the perturbative expansion of the Wilson coefficient
is also $c/z$, which is small for a large $z$. 
This indicates that Wilson coefficients have large perturbative errors at large distance. In Appendix B of Ref.~\cite{Gao:2022iex}, they find that RGR Wilson coefficients can describe some lattice data very well when they choose the starting scale $c/z$ no less than two times $2e^{-\gamma_{E}}/z$. We think one should also consider the starting scale smaller than $2e^{-\gamma_{E}}/z$ to give a reasonable estimation of higher order effects.

\begin{figure}[tbp]
\includegraphics[width=8.5cm]{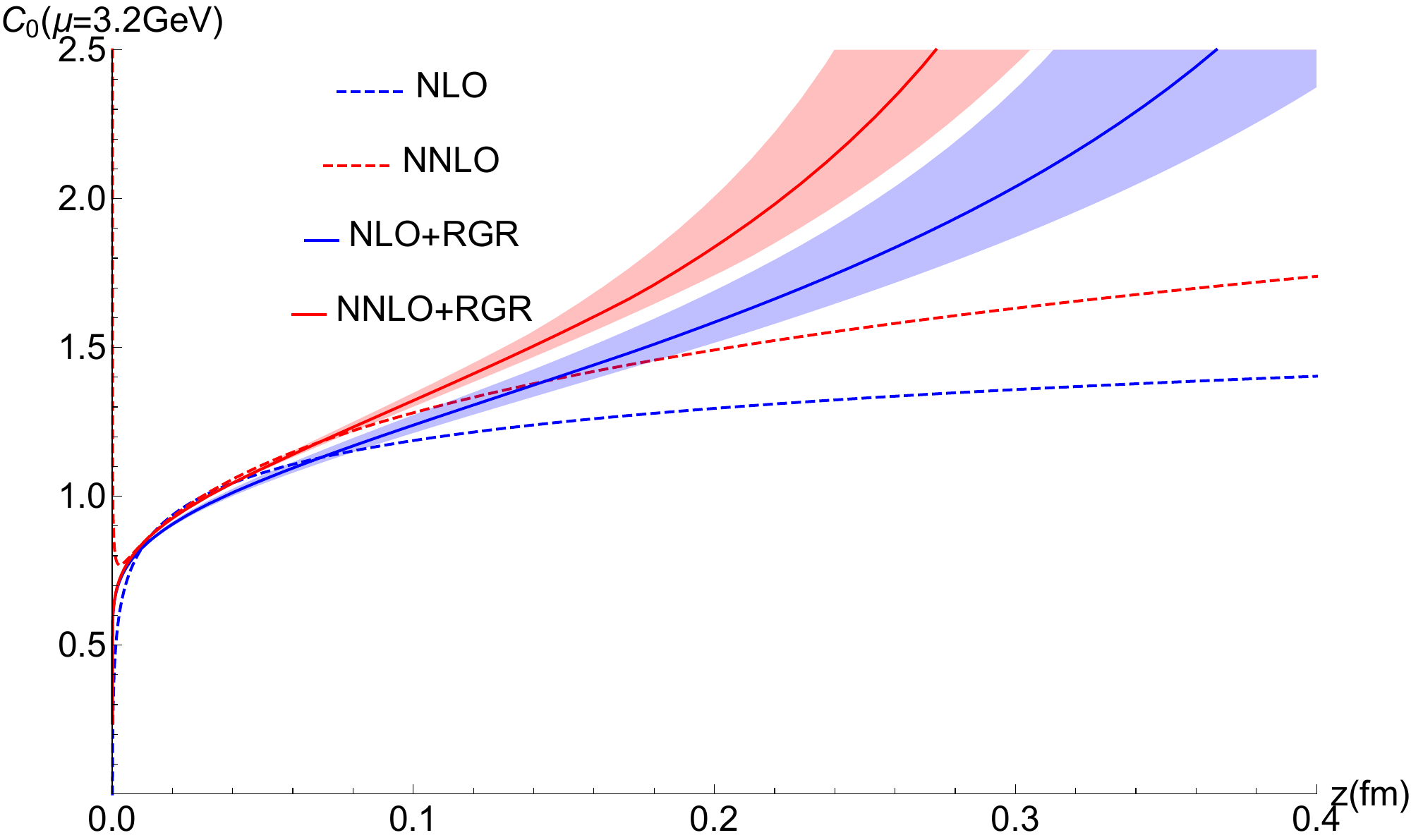}
\caption{The fixed order and RGR Wilson coeffcients $C_{0}$ up to NNLO at $\mu=3.2$ GeV. The bands show the uncertainty from varying $c=(0.8\sim1.2)*2e^{-\gamma_{E}}$.}
\label{fig:C0}
\end{figure}
To see the difference of the fixed-order and resummed Wilson coeffcients, we compare $C^{\overline{\rm MS}}_{0}$ and $C^{\overline{\rm MS}\,{\rm RGR}}_{0}$ at NLO and NNLO at $\mu=3.2$ GeV in Fig.~\ref{fig:C0}. Different types of perturbation series are consistent near $z=0.02 \sim 0.1$ fm. The discrepancy becomes significant for $z > 0.2$~fm between resummed and fixed-order results, as well as between NNLO and NLO. Moreover, the small variation of $c$ causes larger errors at large distance. For $z > 0.4$ fm, the running coupling constant is close to the Landau pole where the perturbation series breaks down, and the OPE analysis is no longer useful.

\subsection{Fitting Moments with Resummed Wilson Coefficients to NNLO}

Combining Eqs.~\eqref{eq:ratio}~\eqref{eq:ratioMSbarfixedorder}~\eqref{eq:OPE}, we have~\cite{Fan:2020nzz,Gao:2020ito}:
\bea\label{eq:ratioOPE}
&&\tilde{h}^{\rm ratio}(z, P_{z}, P_{z}^0) \nonumber\\  &&=\frac{\sum_{N=0}\frac{(-i z P_z)^N}{N!} C^{\overline{\rm MS}}_{N}(\alpha_{s},z^2 \mu^2) \, \langle x^N \rangle(\mu)}{\sum_{N=0}\frac{(-i z P_{z}^0)^N}{N!} C^{\overline{\rm MS}}_{N}(\alpha_{s},z^2 \mu^2) \, \langle x^N \rangle(\mu)}.
\eea
One can use this equation to extract moments from lattice data.

\begin{figure}[htbp]
\includegraphics[width=8.5cm]{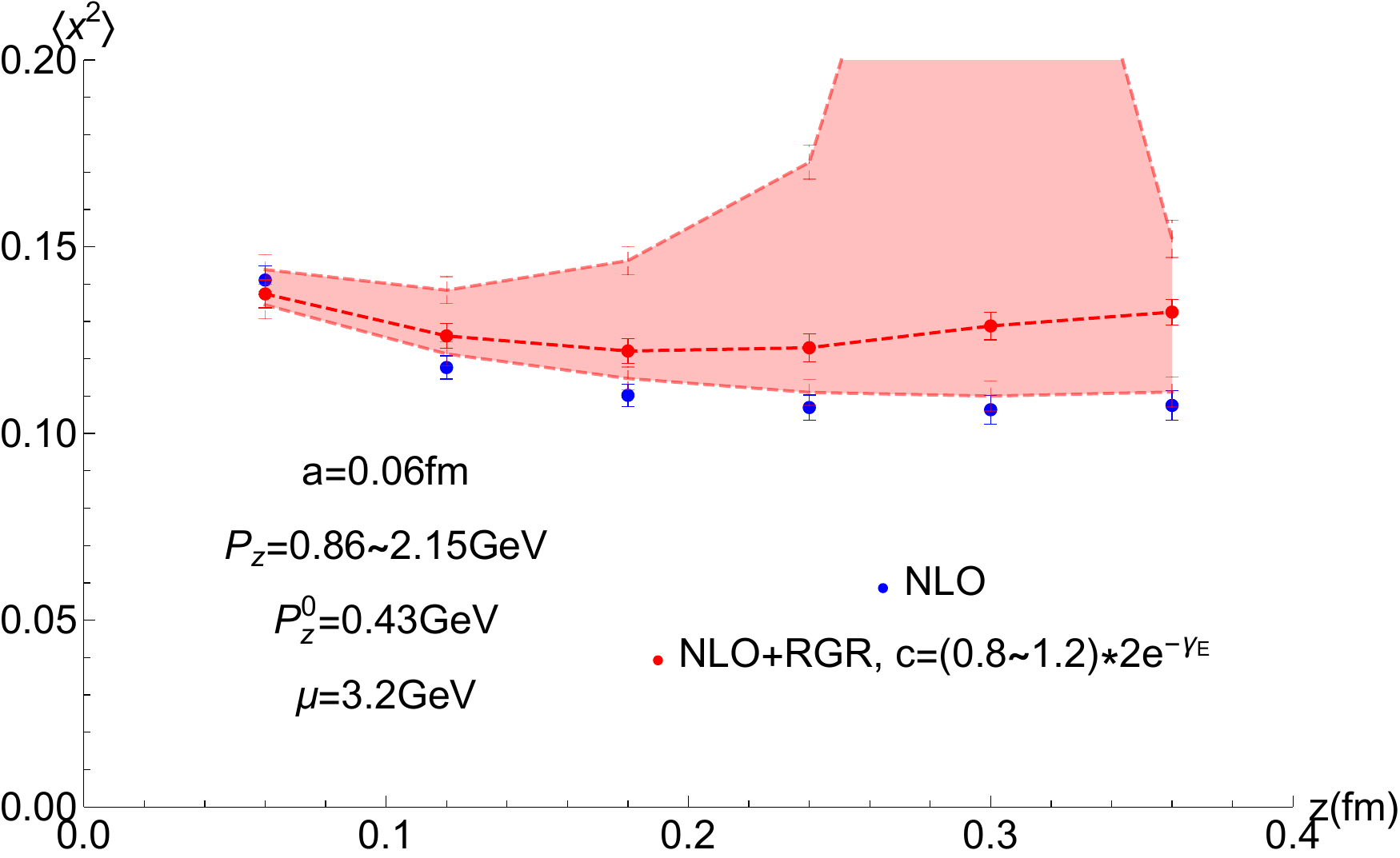}
\caption{The best fit parameter $\langle x^{2} \rangle$ at each single $z$. NLO (blue) and NLO+RGR (red) Wilson coefficients are used at $\mu = 3.2$ GeV. The red band shows the uncertainty from varying $c=(0.8\sim1.2)*2e^{-\gamma_{E}}$.}
\label{fig:x2fixedz}
\end{figure}
Following previous work~\cite{Gao:2020ito},
we fit the $P_z$ dependence at each single $z$ separately. The pion valence PDF is symmetric 
with respect to $x=0$ so it only has even moments. We adopt the following inequalities during the fit: $\langle x^{N+2} \rangle < \langle x^{N} \rangle$, $\langle x^{N+2} \rangle + \langle x^{N-2} \rangle - 2\langle x^{N} \rangle > 0$, and fit up to $\langle x^{6} \rangle$. By using NLO coefficient function, the fit parameter $\langle x^{2} \rangle$ with $P_z = 0.86 \sim 2.15$ GeV at lattice spacing $a$ = 0.06 fm is shown in Fig.~\ref{fig:x2fixedz}. If we just focus on the fixed-order NLO, we see that $\langle x^{2} \rangle$ tends to a stable result as $z$ becomes larger, as observed in~\cite{Gao:2020ito}.
However, when we fit the data with NLO+RGR Wilson coefficients, the variation of $c$ generates an error for $\langle x^{2} \rangle$ and the result at large distance becomes quite uncertain. The result beyond 0.3 fm suffers from large errors. Actually, at $z=0.3$ fm, the starting scale of RGR is 0.59 $\sim$ 0.89 GeV, whose corresponding $\alpha_{s}^{\rm NLO}$ is 0.80 $\sim$ 0.49. $\alpha_{s}^{\rm NLO}=0.80$ is already too large for a perturbation series to converge. So we will only use $z$ up to 0.3 fm in the following analysis in this section.

\begin{figure}[htbp]
\includegraphics[width=8.5cm]{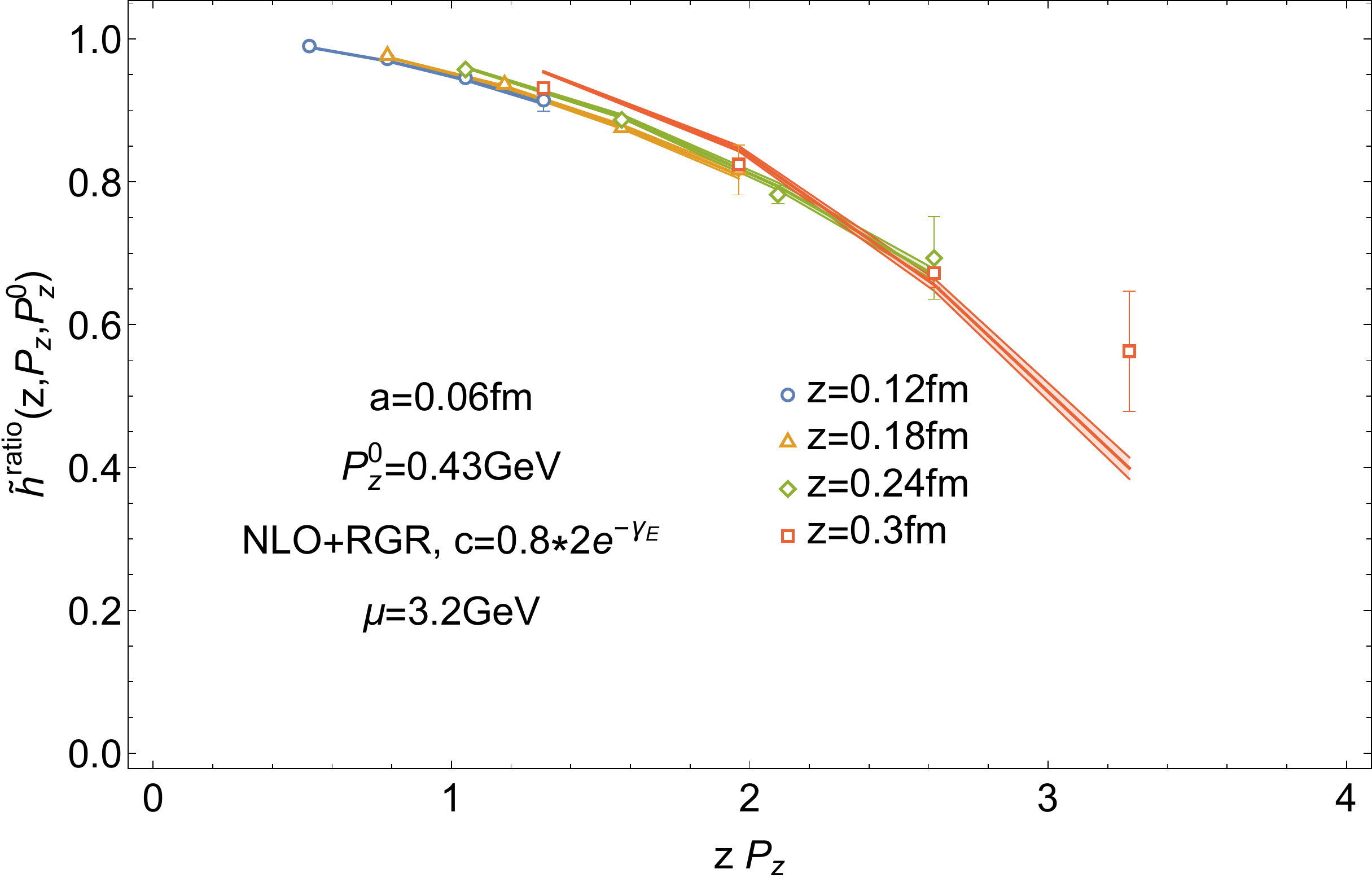}
\includegraphics[width=8.5cm]{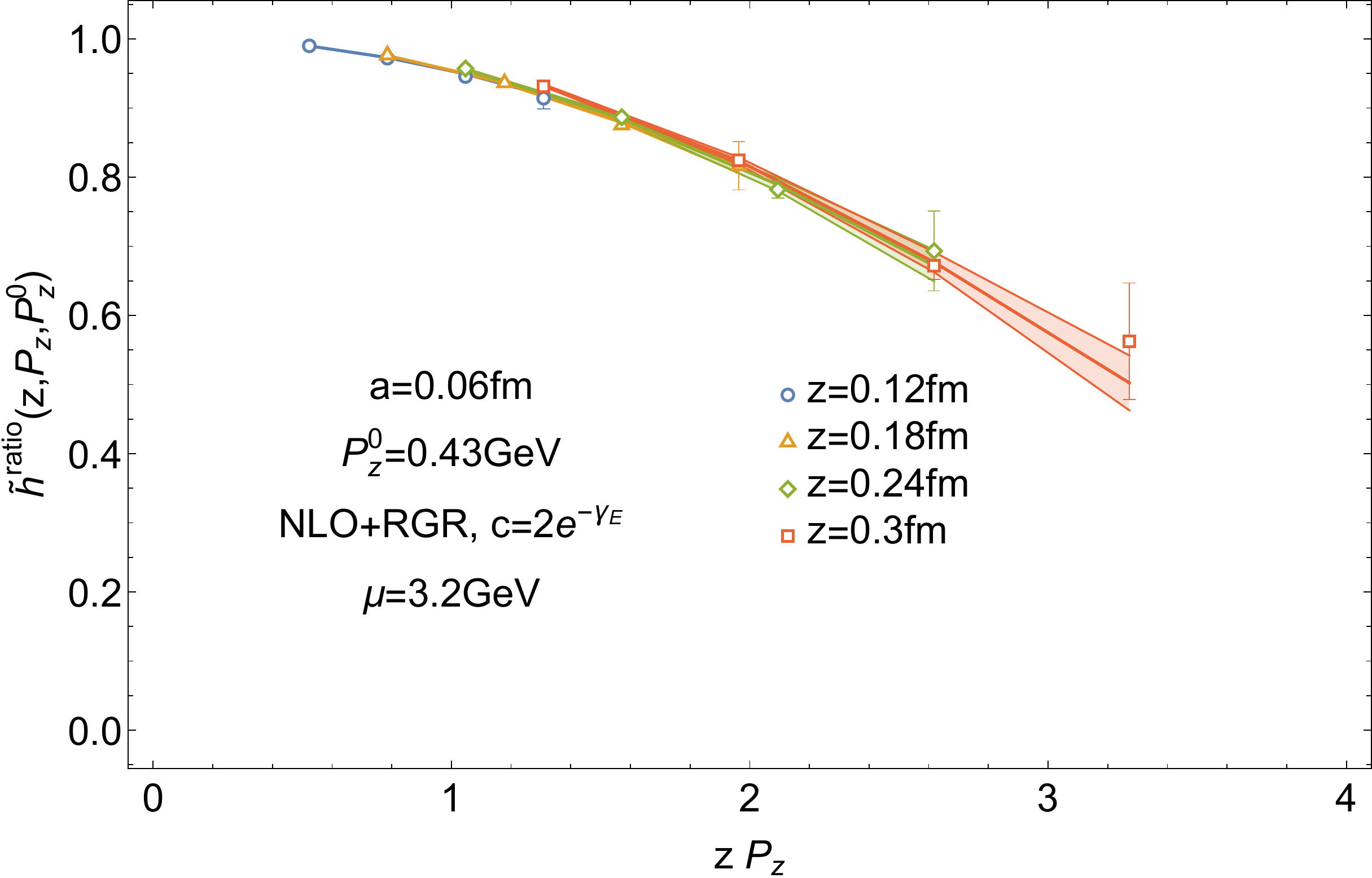}
\includegraphics[width=8.5cm]{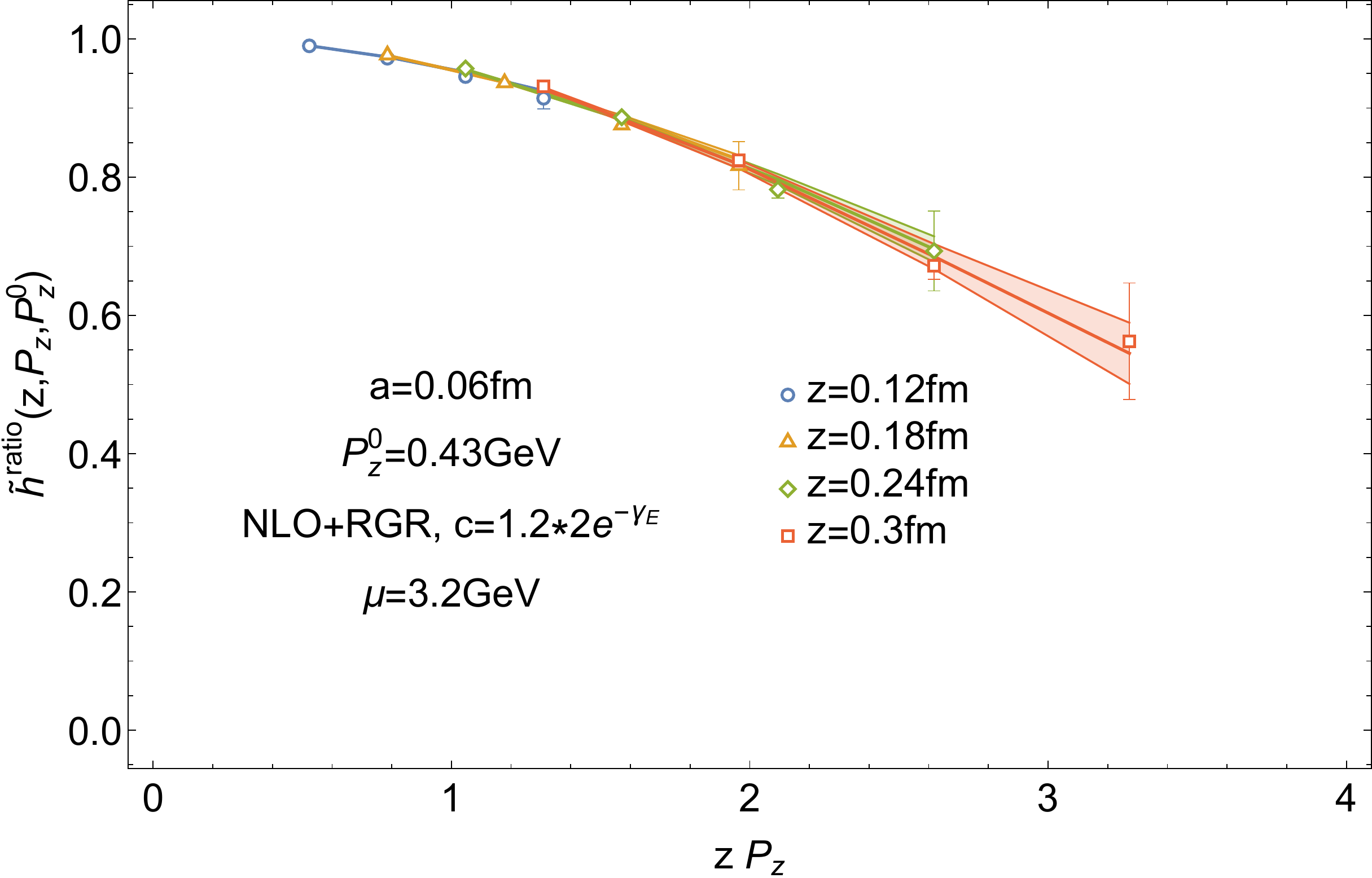}
\caption{A joint fit based on NLO+RGR Wilson coefficients (Eq.~\eqref{eq:WCRGI}) at $\mu=3.2$ GeV. Open dots are data points and bands are fit curves. For $c=0.8*2e^{-\gamma_{E}}$, the best fit $\langle x^{2} \rangle = 0.1588(39)$ and $\langle x^{4} \rangle = 0.106(03)$; For $c=2e^{-\gamma_{E}}$, the best fit $\langle x^{2} \rangle = 0.1251(34)$ and $\langle x^{4} \rangle = 0.029(28)$; For $c=1.2*2e^{-\gamma_{E}}$, the best fit $\langle x^{2} \rangle = 0.1172(32)$ and $\langle x^{4} \rangle = 0.049(16)$.}
\label{fig:FitExampleNLORGI}
\end{figure}

We perform a joint fit with NLO+RGR Wilson coefficient. The joint fit with $P_z = 0.86 \sim 2.15$ GeV and $z = 0.12 \sim 0.3$ fm at lattice spacing $a$ = 0.06 fm at $\mu=3.2$ GeV is shown in Fig.~\ref{fig:FitExampleNLORGI}. We vary $c=(0.8\sim1.2)*2e^{-\gamma_{E}}$ to test the stability of Wilson coefficient and moment fit. The best fit $\langle x^{2} \rangle = 0.1251(34)(221)$ and $\langle x^{4} \rangle = 0.029(28)(40)$, where the second bracket after the center value denotes the systematic error caused by varying $c=(0.8\sim1.2)*2e^{-\gamma_{E}}$. So $\langle x^{4} \rangle$ suffers from large statistical and systematic errors. Note that in the plot of $c=0.8*2e^{-\gamma_{E}}$, the best fit curve does not go through the data point at $z=0.3$ fm.

Finally, we make systematic joint fits on dependence of $P_z$ and $z$ based on NLO, NNLO, NLO+RGR and NNLO+RGR Wilson coefficients at $\mu=3.2$ GeV. 
The best fit parameters $\langle x^{2} \rangle$ and $\langle x^{4} \rangle$ are shown in Table~\ref{tab:fitmoments}. We fit the data at lattice spacing $a=0.06$ fm and $a=0.04$ fm and do a continuum extrapolation $a^2 \rightarrow 0$. The momentum of the matrix element in the denominator is $P_z^{0}=\frac{2\pi}{L a} n_{z}^0$, where we choose $n_{z}^0=1,2$. The fit range is $P_{z}>P_{z}^0$ and $z \in [z_{\rm min},z_{\rm max}]$, where we choose $z_{\rm min}=2a,3a$ and $z_{\rm max} \in [0.24,0.30]$ fm ($z_{\rm max}$ = 0.24, 0.30 fm for $a=0.06$ fm and $z_{\rm max}$ = 0.24, 0.28 fm for $a=0.04$ fm). We vary $c=(0.8\sim1.2)*2e^{-\gamma_{E}}$ for NLO+RGR and NNLO+RGR. The mean value among choosing different $n_{z}^0$, $z_{\rm min}$ and $z_{\rm max}$ is considered as the center value (each combination is equally weighted and we choose $c=2e^{-\gamma_{E}}$ for center value). The standard deviation from choosing different $n_{z}^0$, $z_{\rm min}$, $z_{\rm max}$ and $c$ is considered as the systematic error. After continuum extrapolation, $\langle x^{2} \rangle$ are consistent within the error range among different Wilson coefficients. Both statistical and systematic errors are reasonable for $\langle x^{2} \rangle$. 

However, large systematic errors are observed in $\langle x^{4} \rangle$, especially for NLO+RGR case, which come mostly from varying $c$. Our statistical errors for $\langle x^{4} \rangle$ are larger than~\cite{Gao:2020ito,Gao:2021hxl,Gao:2022iex} because we can only legitimately use $z$ up to 0.3 fm, whereas the previous
works use $z$ up to 0.5$\sim$0.8 fm with fixed-order Wilson coefficients. Thus it appears that the determination of $\langle x^{4} \rangle$ through fitting the lattice correlation function has a large uncertainty.

\begin{table}[tbp]
\begin{tabular}{|c|c|c|}
\hline NLO & $\left\langle x^{2}\right\rangle$  & $\left\langle x^{4}\right\rangle$ \\
\hline $a=0.06 \mathrm{fm}$ & $0.1157(38)(55)$ & $0.0369(126)(145)$ \\
\hline $a=0.04 \mathrm{fm}$ & $0.1062(46)(26)$ & $0.0468(129)(63)$ \\
\hline$a^{2} \rightarrow 0$ & $0.0986(91)(55)$ & $0.0547(250)(145)$ \\
\hline
\hline

\hline NNLO & $\left\langle x^{2}\right\rangle$  & $\left\langle x^{4}\right\rangle$ \\
\hline $a=0.06 \mathrm{fm}$ & $0.1154(39)(56)$ & $0.0388(127)(145)$ \\
\hline $a=0.04 \mathrm{fm}$ & $0.1053(46)(22)$ & $0.0456(130)(45)$ \\
\hline$a^{2} \rightarrow 0$ & $0.0972(91)(56)$ & $0.0510(252)(145)$ \\
\hline
\hline

\hline NLO+RGR & $\left\langle x^{2}\right\rangle$ & $\left\langle x^{4}\right\rangle$ \\
\hline $a=0.06 \mathrm{fm}$ & $0.1273(45)(207)$ & $0.0368(263)(383)$ \\
\hline $a=0.04 \mathrm{fm}$ & $0.1147(51)(122)$ & $0.0539(217)(164)$ \\
\hline$a^{2} \rightarrow 0$ & $0.1046(102)(207)$ & $0.0677(441)(383)$ \\
\hline
\hline

\hline NNLO+RGR & $\left\langle x^{2}\right\rangle$ & $\left\langle x^{4}\right\rangle$ \\
\hline $a=0.06 \mathrm{fm}$ & $0.1073(34)(135)$ & $0.0556(71)(133)$ \\
\hline $a=0.04 \mathrm{fm}$ & $0.0998(43)(79)$ & $0.0521(76)(48)$ \\
\hline$a^{2} \rightarrow 0$ & $0.0938(85)(135)$ & $0.0494(147)(133)$ \\
\hline

\end{tabular}
\caption{The best fit parameters $\langle x^{2} \rangle$ and $\langle x^{4} \rangle$ based on NLO, NNLO, NLO+RGR and NNLO+RGR Wilson coefficients at $\mu=3.2$ GeV. The first bracket after the center value denotes the statistical error. The standard deviation from choosing different $P_{z}^0$, $z_{\rm min}$, $z_{\rm max}$ and $c$ is considered as the systematic error, which is shown in the second bracket after the center value.}
\label{tab:fitmoments}
\end{table}

\section{Summary and Outlook}\label{sec:summary}

In this work, we study the effects of small-momentum large logarithm resummation in the perturbative matching of LaMET. The resummed matching improves the prediction accuracy at intermediate $2 x P_z$, and at the same time, clearly indicates that the perturbative matching is unreliable at small $2 x P_z$ (e.g. $2xP_z<0.7$~GeV). This is consistent with power counting in the large-momentum expansion. Besides, we test the RGR effect in OPE approach and show that Wilson coefficients cannot be calculated in perturbation theory at large $z$ (e.g. $z>0.3$~fm). These observations are consistent with each other, as small $x$ region in momentum space conjugates to large $z$ region in coordinate space. 

Thus, to calculate small-$x$ parton physics, one has to 
go to larger hadron momentum than presently available. With increased computational power,
it may be possible to reach $P_z\sim$ 5~GeV, if so one can calculate
on lattice partons at $x\sim 0.1$. To calculate, $x\sim 0.01$, one
has to reach $P_z\sim $ 50~GeV, which cannot be done without a breakthrough
in lattice technology or reformulation of the LaMET formalism. 

A related question is about large-$x$ PDFs. As $x\to 1$, a new soft scale
appears, $(1-x)P_z$, which is the momentum of the hadron remnant. Here again, 
large-logarithms appear due to incomplete cancellation of infrared physics. 
They can be re-summed to produce more reliable predictions in the large-$x$ region. 
At the same time, however, it also exposes the breakdown of LaMET
in the large $x$ region, where the higher-twist corrections again
become important. We hope to address this in a future publication.



\section*{Acknowledgement}
We thank the BNL/ANL lattice collaboration for sharing their data for this work. We also thank Yong Zhao and Peter Petreczky for useful discussions, particularly about the
moment fits. While preparing for this draft, we are noticed that the two-loop hybrid scheme matching kernel is also derived in Ref.~\cite{Zhao:2022xxx}. This research is supported by the U.S. Department of Energy, Office of Science, Office of Nuclear Physics, under contract number DE-SC0020682. Y.S. is partially supported by the U.S.~Department of Energy, Office of Science, Office of Nuclear Physics, contract no.~DE-AC02-06CH11357. F. Yao and J.-H. Zhang are supported by the National Natural Science Foundation of China under grants No. 11975051 and No. 12061131006. J. Holligan is partially supported by the Center for Frontiers in Nuclear Science at Stony Brook University.

\onecolumngrid
\appendix
\section{Hybrid Scheme Matching Kernel in Quasi-PDF Approach}\label{app:hybridkernel}
We relate the hybrid scheme matching kernel with the ratio scheme or $\overline{\rm MS}$ scheme in coordinate space, and double Fourier transform it to momentum space. The matching kernel here is for unpolarized, flavor non-singlet and $\gamma^{t}$ quasi-PDF matrix element.

\subsection{Notation and Convention}
The factorization in coordinate space in the hybrid scheme:
\bea
\tilde{h}(z, P_{z})= &&\int_{-1}^{1} d\nu Z(\nu,|z|,\mu) h(\nu\lambda,\mu) + {\cal O}(z^2 \Lambda_{\rm QCD}^2), 
\eea
where $\lambda = z P_z$. The factorization in momentum space in hybrid scheme:
\bea
\tilde{f}(x, P_{z}) = \int_{-1}^{1} \frac{dy}{\left| y \right|} C\left(\frac{x}{y},\frac{\mu}{|x|P_{z}}\right) f(y, \mu) + {\cal O}\left[\frac{\Lambda_{\rm QCD}^2}{x^2 P_{z}^2}, \frac{\Lambda_{\rm QCD}^2}{(1-x)^2 P_{z}^2}\right],
\eea
where ${\cal O}\left[\frac{\Lambda_{\rm QCD}^2}{x^2 P_{z}^2}, \frac{\Lambda_{\rm QCD}^2}{(1-x)^2 P_{z}^2}\right]$ denotes the power corrections. According to Ref.~\cite{Ji:2020brr,Ji:2020ect}, the LaMET factorization requires both the active quark and spectator to be collinear, which means $x P_z \gg \Lambda_{\rm QCD}$ and $(1-x)P_z \gg \Lambda_{\rm QCD}$. Therefore, by dimensional analysis, 
the power corrections are controlled by small parameters 
$\frac{\Lambda_{\rm QCD}}{x P_{z}}$, $\frac{\Lambda_{\rm QCD}}{(1-x) P_{z}}$. If quasi-PDF operators have no linear divergence or if they do but working in dimensional regularization, the power
corrections start from square powers of these parameters due to symmetry reasons. The Fourier transformations:
\bea
\tilde{f}(x, P_{z}) = P_z \int_{-\infty}^{+\infty} \frac{d z}{2 \pi} e^{i x P_z z} \tilde{h}(z, P_{z}),
\eea
\bea
f(x,\mu) = \int_{-\infty}^{+\infty} \frac{d \lambda}{2 \pi} e^{i x \lambda} h(\lambda, \mu),
\eea
where $\lambda = z P_z$. So the double Fourier transformation is:
\bea\label{eq:doubleFT2}
C\left(\xi, \frac{\mu}{|x| P_{z}}\right) = \int_{-\infty}^{\infty} \frac{d \nu}{2 \pi} e^{i \nu \xi} \int_{-1}^{1} d u e^{-i u \nu} Z\left(u,\left|\frac{\nu}{y P_{z}}\right|, \mu\right),
\eea
where $\xi=x/y$. The relation between ratio scheme matching kernel $Z^{\rm ratio}$ and the $\overline{\rm MS}$ scheme matching kernel $Z^{\overline{\rm MS}}$ in coordinate space is
\bea\label{eq:ratioMSbar}
Z^{\rm ratio}\left(u,z^{2} \mu^{2}\right)=\frac{Z^{\overline{\rm MS}}\left(u,z^{2} \mu^{2}\right)}{\tilde{h}^{\overline{\rm MS}}(z, \mu, 0)},
\eea
where $\tilde{h}^{\overline{\rm MS}}(z, \mu, 0)$ is the perturbative zero-momentum matrix element $\langle P_z=0 | \bar{\psi}(z) \gamma^{t} U(z,0) \psi (0) |P_z=0 \rangle$ in $\overline{\rm MS}$ scheme. In principle, any finite-momentum matrix element can be used
to cancel the linear divergence. However, the zero-momentum matrix element contains only non-perturbative normalization factor, with infrared finite $z$-dependence calculable entirely in perturbation theory. Diving by matrix elements at other momenta will change the infrared structure, thus the factorization may break down. $\tilde{h}^{\overline{\rm MS}}(z, \mu, 0)$ is calculated up to NNLO in literature~\cite{Li:2020xml}:
\bea\label{eq:MSbarpert}
&&\tilde{h}^{\overline{\rm MS}}(z, \mu, 0)= \tilde{h}^{\overline{\rm MS}(0)}(z, \mu, 0)+\tilde{h}^{\overline{\rm MS}(1)}(z, \mu, 0)+\tilde{h}^{\overline{\rm MS}(2)}(z, \mu, 0)\nonumber\\
&&=1+\frac{\alpha_{s}C_{F}}{2 \pi}\left(\frac{3}{2}\ln\left(\frac{z^{2}\mu^{2}e^{2\gamma_E}}{4}\right)+\frac{5}{2}\right) \nonumber\\
&&+\left(\frac{\alpha_s}{2\pi}\right)^2\left[\left(\frac{11 C_A C_F}{8}-\frac{C_F n_f T_F}{2}+\frac{9 C_F^2}{8}\right) \ln^2 \left(\frac{z^{2}\mu^{2}e^{2\gamma_E}}{4}\right) \right.\nonumber\\
&&\left.+\left(\left(\frac{53}{8}-\frac{\pi ^2}{6}\right) C_A C_F+\left(\frac{25}{8}+\frac{2\pi ^2}{3}\right) C_F^2 -\frac{5 C_F n_f T_F}{2}\right) \ln \left(\frac{z^{2}\mu^{2}e^{2\gamma_E}}{4}\right) \right. \nonumber\\
&&\left.+2 \left(-4 \zeta (3)+\frac{223}{192} +\frac{\pi^2}{9}\right) C_F^2 +2\left(\zeta (3)+\frac{4877}{576}-\frac{5\pi ^2}{24}\right) C_A C_F-\frac{469 C_F n_f T_F}{72}\right],
\eea
where we use the superscripts (0), (1), (2) to denote terms at ${\cal O}(1)$, ${\cal O}(\alpha_s)$ and ${\cal O}(\alpha_s^2)$ separately. $T_F=\frac{1}{2}$ is a conventional normalization factor for SU(3) generators in the fundamental representation. $C_{F}=\frac{4}{3}$ and $C_{A}=3$ are the quadratic Casimir operators for SU(3) in the fundamental and adjoint representation separately. $n_f$ is the number of quark flavors. $Z^{\rm ratio}\left(u,z^{2} \mu^{2}\right)$ up to NLO~\cite{Radyushkin:2018cvn,Zhang:2018ggy,Izubuchi:2018srq} is
\bea
Z^{\rm ratio}\left(u,z^{2} \mu^{2}\right)= \delta(1-u)+\frac{\alpha_{s} C_{F}}{2 \pi}\left[-\frac{1+u^{2}}{1-u} \ln\left( \frac{ z^2 \mu^2 e^{2 \gamma_{E}} }{4} \right) +\frac{1-4 u+u^{2}-4 \ln (1-u)}{1-u}\right]_{+(1)}^{[0,1]} \theta(u)\theta(1-u),
\eea
and $Z^{\overline{\rm MS}}\left(u,z^{2} \mu^{2}\right)$ up to NLO is:
\bea\label{eq:MSbarkernelNLOcoor}
&&Z^{\overline{\rm MS}}(u,z^2 \mu^2)=
Z^{\rm ratio(0)}\left(u,z^{2} \mu^{2}\right) + Z^{\rm ratio(1)}\left(u,z^{2} \mu^{2}\right) + \delta(1-u)\frac{\alpha_{s} C_{F}}{2 \pi}\left(\frac{3}{2}\ln\left(\frac{z^{2}\mu^{2}e^{2\gamma_E}}{4}\right)+\frac{5}{2}\right),
\eea
where we use the superscripts (0) and (1) to denote terms at ${\cal O}(1)$ and ${\cal O}(\alpha_s)$ separately.

The relation between hybrid scheme matching kernel $Z$ and ratio scheme matching kernel $Z^{\rm ratio}$ in coordinate space is (it follows the same convention as how we renormalize the lattice matrix element in Eq.~\eqref{eq:RenonpertHybrid1})
\bea\label{eq:hybridratio}
Z(u,|z|,\mu)&&=Z^{\rm ratio}(u,z^2\mu^2)\theta(z_s-|z|) + \frac{Z^{\overline{\rm MS}}(u,z^2\mu^2)}{\tilde{h}^{\overline{\rm MS}}(z_s, \mu, 0)}\theta(|z|-z_s) \nonumber\\
&&= Z^{\rm ratio}(u,z^2\mu^2)\left[1 + \left(\frac{\tilde{h}^{\overline{\rm MS}}(z, \mu, 0)}{\tilde{h}^{\overline{\rm MS}}(z_s, \mu, 0)}-1\right)\theta(|z|-z_s)\right].
\eea

We would like to expand Eq.~\eqref{eq:hybridratio} at ${\cal O}(1)$, ${\cal O}(\alpha_s)$ and ${\cal O}(\alpha_s^2)$:
\bea
Z(u,|z|,\mu) = Z^{(0)}(u,|z|,\mu) + Z^{(1)}(u,|z|,\mu) + Z^{(2)}(u,|z|,\mu),
\eea
and get the hybrid scheme matching kernel in momentum space through double Fourier transformation Eq.~\eqref{eq:doubleFT2} order by order:
\bea
C\left(\xi, \frac{\mu}{|x| P_{z}}\right)=C^{(0)}\left(\xi, \frac{\mu}{|x| P_{z}}\right)+C^{(1)}\left(\xi, \frac{\mu}{|x| P_{z}}\right)+C^{(2)}\left(\xi, \frac{\mu}{|x| P_{z}}\right).
\eea

\subsection{LO and NLO Matching Kernel}
At LO, the hybrid scheme matching kernel in coordinate space is
$$Z^{(0)}(u,|z|,\mu)=\delta(1-u).$$
We double Fourier transform it based on Eq.~\eqref{eq:doubleFT2} to get the hybrid scheme matching kernel in momentum space at LO, 
$$C^{(0)}\left(\xi, \frac{\mu}{|x| P_{z}}\right)=\delta(1-\xi).$$

At NLO, the hybrid scheme matching kernel in coordinate space is
\bea
Z^{(1)}(u,|z|,\mu)&&=Z^{\rm ratio(1)}\left(u,z^{2} \mu^{2}\right) + Z^{\rm ratio(0)}\left(u,z^{2} \mu^{2}\right)(\tilde{h}^{\overline{\rm MS}(1)}(z, \mu, 0)-\tilde{h}^{\overline{\rm MS}(1)}(z_s, \mu, 0))\theta(|z|-z_s) \nonumber\\
&&=Z^{\rm ratio(1)}\left(u,z^{2} \mu^{2}\right)+\delta(1-u)\frac{\alpha_s C_{F}}{2 \pi}\frac{3}{2}\ln\left(\frac{z^{2}}{z_s^{2}}\right)\theta(|z|-z_s).
\eea
We double Fourier transform it based on Eq.~\eqref{eq:doubleFT2} and get the hybrid scheme matching kernel in momentum space at NLO~\cite{Chou:2022drv}:
\bea\label{eq:hybridkernelNLO}
C^{(1)}\left(\xi,\frac{\mu}{|x|P_{z}}\right) = C^{\rm ratio(1)}\left(\xi, \frac{\mu}{|x| P_{z}}\right) + \frac{\alpha_s C_{F}}{2 \pi}\frac{3}{2} \left[-\frac{1}{|1-\xi|}+\frac{2 {\rm Si}[(1-\xi)|y| z_s P_z]}{\pi (1-\xi)} \right]_{+(1)}^{[-\infty,\infty]},
\eea
where $2 {\rm Si}[(1-\xi)|y| z_s P_z]$ is the sine integral function and $C^{\rm ratio(1)}\left(\xi, \frac{\mu}{|x| P_{z}}\right)$ is the ratio scheme matching kernel in momentum space at NLO~\cite{Izubuchi:2018srq}:
\bea\label{eq:ratiokernelNLO}
C^{\rm ratio(1)}\left(\xi, \frac{\mu}{|x| P_{z}}\right)= \frac{\alpha_{s} C_{F}}{2 \pi} \begin{cases}\left(\frac{1+\xi^{2}}{1-\xi} \ln \frac{\xi}{\xi-1}+1-\frac{3}{2(1-\xi)}\right)_{+(1)}^{[1, \infty]} & \xi>1 \\ \left(\frac{1+\xi^{2}}{1-\xi}\left[-\ln \frac{\mu^{2}}{4 x^{2} P_{z}^{2}}+\ln (\frac{1-\xi}{\xi})-1\right]+1+\frac{3}{2(1-\xi)}\right)_{+(1)}^{[0,1]} & 0<\xi<1 \\ \left(-\frac{1+\xi^{2}}{1-\xi} \ln \frac{-\xi}{1-\xi}-1+\frac{3}{2(1-\xi)}\right)_{+(1)}^{[-\infty, 0]} & \xi<0 \end{cases}
\eea

\subsection{NNLO Matching Kernel}
At NNLO, the hybrid scheme matching kernel in coordinate space is
\bea\label{eq:hybrid2}
&&Z^{(2)}(u,|z|,\mu)=Z^{\rm ratio(2)}\left(u,z^{2} \mu^{2}\right)+Z^{\rm ratio(1)}\left(u,z^{2} \mu^{2}\right)[\tilde{h}^{\overline{\rm MS}(1)}(z, \mu, 0)-\tilde{h}^{\overline{\rm MS}(1)}(z_s, \mu, 0)]\theta(|z|-z_s) \nonumber\\ 
&&+ Z^{\rm ratio(0)}\left(u,z^{2} \mu^{2}\right)[\tilde{h}^{\overline{\rm MS}(2)}(z, \mu, 0)-\tilde{h}^{\overline{\rm MS}(2)}(z_s, \mu, 0)-\tilde{h}^{\overline{\rm MS}(1)}(z_s, \mu, 0)(\tilde{h}^{\overline{\rm MS}(1)}(z, \mu, 0)-\tilde{h}^{\overline{\rm MS}(1)}(z_s, \mu, 0))]\theta(|z|-z_s) \nonumber\\ 
\eea
Note that it is hard to find the analytical result of double Fourier transformation on $Z^{\rm ratio(1)}\left(u,z^{2} \mu^{2}\right)[\tilde{h}^{\overline{\rm MS}(1)}(z, \mu, 0)-\tilde{h}^{\overline{\rm MS}(1)}(z_s, \mu, 0)]\theta(|z|-z_s)$ and it is also hard to do it numerically since it contains a singularity for $\xi \rightarrow 1$. However, one can make use of the following relations:
$$
Z^{\rm ratio(2)}\left(u,z^{2} \mu^{2}\right)=Z^{\overline{\rm MS}(2)}\left(u,z^{2} \mu^{2}\right)-\delta(1-u)\tilde{h}^{\overline{\rm MS}(2)}(z, \mu, 0)-Z^{\rm ratio(1)}\left(u,z^{2} \mu^{2}\right)\tilde{h}^{\overline{\rm MS}(1)}(z, \mu, 0),
$$
and
$$
Z^{\rm ratio(1)}\left(u,z^{2} \mu^{2}\right)[\tilde{h}^{\overline{\rm MS}(1)}(z, \mu, 0)-\tilde{h}^{\overline{\rm MS}(1)}(z_s, \mu, 0)]\theta(|z|-z_s)=Z^{\rm ratio(1)}\left(u,z^{2} \mu^{2}\right)[\tilde{h}^{\overline{\rm MS}(1)}(z, \mu, 0)-\tilde{h}^{\overline{\rm MS}(1)}(z_s, \mu, 0)][1-\theta(z_s-|z|)].
$$
Substituting the above two relations into Eq.~\eqref{eq:hybrid2}, we have
\bea\label{eq:hybrid2new}
&&Z^{(2)}(u,|z|,\mu)=Z^{\overline{\rm MS}(2)}\left(u,z^{2} \mu^{2}\right)-\delta(1-u)\tilde{h}^{\overline{\rm MS}(2)}(z, \mu, 0) \nonumber\\ 
&&+ \delta(1-u)[\tilde{h}^{\overline{\rm MS}(2)}(z, \mu, 0)-\tilde{h}^{\overline{\rm MS}(2)}(z_s, \mu, 0)-\tilde{h}^{\overline{\rm MS}(1)}(z_s, \mu, 0)(\tilde{h}^{\overline{\rm MS}(1)}(z, \mu, 0)-\tilde{h}^{\overline{\rm MS}(1)}(z_s, \mu, 0))]\theta(|z|-z_s)\nonumber\\ 
&&-Z^{\rm ratio(1)}\left(u,z^{2} \mu^{2}\right)\tilde{h}^{\overline{\rm MS}(1)}(z_s, \mu, 0)-Z^{\rm ratio(1)}\left(u,z^{2} \mu^{2}\right)[\tilde{h}^{\overline{\rm MS}(1)}(z, \mu, 0)-\tilde{h}^{\overline{\rm MS}(1)}(z_s, \mu, 0)]\theta(z_s-|z|),
\eea
where we can find analytical results of double Fourier transformation on all these terms except for the last one $Z^{\rm ratio(1)}\left(u,z^{2} \mu^{2}\right)[\tilde{h}^{\overline{\rm MS}(1)}(z, \mu, 0)-\tilde{h}^{\overline{\rm MS}(1)}(z_s, \mu, 0)]\theta(z_s-|z|)$. The result of double Fourier transformation on $Z^{\rm ratio(1)}\left(u,z^{2} \mu^{2}\right)[\tilde{h}^{\overline{\rm MS}(1)}(z, \mu, 0)-\tilde{h}^{\overline{\rm MS}(1)}(z_s, \mu, 0)]\theta(z_s-|z|)$ is a smooth function with respect to $\xi$ and we can calculate it numerically. 

We double Fourier transform Eq.~\eqref{eq:hybrid2new} based on Eq.~\eqref{eq:doubleFT2} and get the hybrid scheme matching kernel in momentum space at NNLO:
\bea\label{eq:hybridkernelNNLO}
&&C^{(2)}\left(\xi,\frac{\mu}{|x|P_{z}}\right) = \Bigg\{ C^{\overline{\rm MS}(2)}\left(\xi,\frac{\mu}{|x|P_{z}}\right) -\left(\frac{\alpha_s}{2\pi}\right)^2 \left[\left(\frac{11 C_A C_F}{8}-\frac{C_F n_f T_F}{2}+\frac{9 C_F^2}{8}\right)\frac{2\ln\left(\frac{4 x^2 P_z^2 (1-\xi)^2}{\mu^2 \xi^2} \right)}{|1-\xi|} \right.\nonumber\\
&&\left.+\left(\left(\frac{53}{8}-\frac{\pi ^2}{6}\right) C_A C_F+\left(\frac{25}{8}+\frac{2\pi ^2}{3}\right) C_F^2 -\frac{5 C_F n_f T_F}{2}\right)\left(-\frac{1}{|1-\xi|}\right) \right] \nonumber\\
&&+\left(\frac{\alpha_s}{2\pi}\right)^2 \left(\frac{11 C_A C_F}{8}-\frac{C_F n_f T_F}{2}+\frac{9 C_F^2}{8}\right)\left[\frac{2}{|1-\xi|}\ln \left(y^2 P_z^2 z_s^2 e^{2 \gamma_{E}} (1-\xi)^2 \right) \right. \nonumber\\ 
&&\left. -\frac{4|y|P_z z_s}{\pi}\left( { }_{3} F_{3}\left(1,1,1 ; 2,2,2 ; i\left(\xi-1\right) |y|P_z z_s\right)+{ }_{3} F_{3}\left(1,1,1 ; 2,2,2 ;-i\left(\xi-1\right) |y|P_z z_s\right) \right) \right. \nonumber\\ 
&&\left. +2\left(-\frac{1}{|1-\xi|}+\frac{2 {\rm Si}[(1-\xi)|y| z_s P_z]}{\pi (1-\xi)} \right)\ln \left(\frac{z_s^{2}\mu^{2}e^{2\gamma_E}}{4}\right) \right] \nonumber\\ 
&&+\left(\frac{\alpha_s}{2\pi}\right)^2 \left(\left(\frac{53}{8}-\frac{\pi ^2}{6}\right) C_A C_F+\left(\frac{25}{8}+\frac{2\pi ^2}{3}\right) C_F^2 -\frac{5 C_F n_f T_F}{2}\right)\left(-\frac{1}{|1-\xi|}+\frac{2 {\rm Si}[(1-\xi)|y| z_s P_z]}{\pi (1-\xi)} \right) \nonumber\\ 
&&-\left(\frac{\alpha_s}{2\pi}\right)^2 C_F^2 \left(\frac{3}{2}\ln\left(\frac{z_s^{2}\mu^{2}e^{2\gamma_E}}{4}\right)+\frac{5}{2}\right)\frac{3}{2}\left(-\frac{1}{|1-\xi|}+\frac{2 {\rm Si}[(1-\xi)|y| z_s P_z]}{\pi (1-\xi)} \right) \nonumber\\ 
&&-C^{\rm ratio(1)}\left(\xi,\frac{\mu}{|x|P_z}\right)\frac{\alpha_{s}C_{F}}{2 \pi}\left(\frac{3}{2}\ln\left(\frac{z_s^{2}\mu^{2}e^{2\gamma_E}}{4}\right)+\frac{5}{2}\right) \nonumber\\
&&-\int_{-\infty}^{\infty} \frac{d \nu}{2 \pi} e^{i \nu \xi} \int_{-1}^{1} d u e^{-i u \nu} Z^{\rm ratio(1)}\left(u,\left(\frac{\nu}{y P_{z}}\right)^{2} \mu^{2}\right)\frac{\alpha_s C_{F}}{2 \pi}\frac{3}{2}\ln\left[\left(\frac{\nu}{y P_{z}z_s}\right)^{2}\right]\theta\left(z_s-\left|\frac{\nu}{y P_{z}}\right|\right) \Bigg\}_{+(1)}^{[-\infty,+\infty]},
\eea
where $C^{\overline{\rm MS}(2)}$ is the two-loop matching kernel for flavor non-singlet case in $\overline{\rm MS}$ scheme, which can be found in~\cite{Li:2020xml,Chen:2020ody}. ${ }_{3} F_{3}$ is the hypergeometric function. Note that we drop away all the terms proportional to $\delta(1-\xi)$ during the deduction and put an overall plus function in the end to guarantee the current conservation.

\section{NNLO Wilson Coefficient and Anomalous Dimension in OPE }\label{app:Wilcoe}

The Wilson coefficient $C^{\overline{\rm MS}}_{N}(\alpha_{s},z^2 \mu^2)$ is the moment of coordinate space matching kernel $Z^{\overline{\rm MS}}(\nu,z^2 \mu^2)$:
\bea
C^{\overline{\rm MS}}_{N}(\alpha_{s},z^2 \mu^2)=\int_{-1}^{1} d\nu \, \nu^N Z^{\overline{\rm MS}}(\nu,z^2 \mu^2).
\eea
We present Wilson coefficients order by order
\bea
C^{\overline{\rm MS}}_{N}(\alpha_{s},z^2 \mu^2)=1+C^{\overline{\rm MS}(1)}_{N}(\alpha_{s},z^2 \mu^2)+C^{\overline{\rm MS}(2)}_{N}(\alpha_{s},z^2 \mu^2).
\eea

At NLO, we have~\cite{Gao:2021hxl}:
\bea\label{eq:WilcoeNLO}
C^{\overline{\rm MS}(1)}_{N}(\alpha_{s},z^2 \mu^2)=\frac{\alpha_{s} C_{F}}{2 \pi}\left[\left(\frac{3+2 N}{2+3 N+N^{2}}+2 H_{N}\right) \ln \left(\frac{z^{2}\mu^{2}e^{2\gamma_E}}{4}\right)+\frac{5+2 N}{2+3 N+N^{2}}+2\left(1-H_{N}\right) H_{N}-2 H_{N}^{(2)}\right]
\eea
where $H_{N}=\sum_{i=1}^{N}1/i$ and $H_{N}^{(2)}=\sum_{i=1}^{N}1/i^2$.

At NNLO, we get the Wilson coefficients for flavor non-singlet case (flavor non-singlet is the same as valence for pion) based on~\cite{Li:2020xml}. We use their supplemental material on prl page including a Mathematica package. We find an analytical form for general $N$ for the term proportional to $\ln^2 \left(\frac{z^{2}\mu^{2}e^{2\gamma_E}}{4}\right)$:
\bea
&&C^{\overline{\rm MS}(2)}_{N}(\alpha_{s},z^2 \mu^2)=\left(\frac{\alpha_s}{2\pi}\right)^2\left[\frac{11 C_A C_F}{8}-\frac{C_F n_f T_F}{2}+\frac{9 C_F^2}{8} +  C_A C_F \frac{11}{24} \left(4 H_N+\frac{2}{N+1}+\frac{2}{N+2}-3\right) \right. \nonumber\\
&&\left. + C_F^2 \left(\frac{2 (2N+3) H_{N+1}}{N^2+3N+2}+2 \left(H_N\right)^2-\frac{N (N (9 N (N+6)+133)+172)+96}{8\left(N^2+3N+2\right)^2}\right)\right. \nonumber\\
&&\left.+C_F n_f T_F \frac{1}{6} \left(\frac{N (3N+5)}{(N+1) (N+2)}-4 H_N\right) \right]\ln^2 \left(\frac{z^{2}\mu^{2}e^{2\gamma_E}}{4}\right) + ...
\eea
We present the numerical results for Wilson coefficients at NNLO up to $N=12$:
\bea\label{eq:WilcoeNNLO}
&&C^{\overline{\rm MS}(2)}_{0}(\alpha_{s},z^2 \mu^2)=\left(\frac{\alpha_s}{2\pi}\right)^2[L^2 \left(7.5\, -0.333333 n_f\right)+L \left(37.1731\, -1.66667 n_f\right)-4.34259 n_f+51.836] \nonumber\\
&&C^{\overline{\rm MS}(2)}_{1}(\alpha_{s},z^2 \mu^2)=\left(\frac{\alpha_s}{2\pi}\right)^2[L^2 \left(17.5247\, -0.62963 n_f\right)+L \left(13.9468\, -0.975309 n_f\right)-6.56481 n_f+63.8866] \nonumber\\
&&C^{\overline{\rm MS}(2)}_{2}(\alpha_{s},z^2 \mu^2)=\left(\frac{\alpha_s}{2\pi}\right)^2[L^2 \left(24.5525\, -0.796296 n_f\right)+L \left(-0.391975 n_f-14.6427\right)-8.33835 n_f+95.0155] \nonumber\\
&&C^{\overline{\rm MS}(2)}_{3}(\alpha_{s},z^2 \mu^2)=\left(\frac{\alpha_s}{2\pi}\right)^2[L^2 \left(30.1584\, -0.914815 n_f\right)+L \left(0.129506 n_f-42.8518\right)-9.89856 n_f+132.071] \nonumber\\
&&C^{\overline{\rm MS}(2)}_{4}(\alpha_{s},z^2 \mu^2)=\left(\frac{\alpha_s}{2\pi}\right)^2[L^2 \left(34.8899\, -1.00741 n_f\right)+L \left(0.601728 n_f-69.9135\right)-11.3178 n_f+171.79] \nonumber\\
&&C^{\overline{\rm MS}(2)}_{5}(\alpha_{s},z^2 \mu^2)=\left(\frac{\alpha_s}{2\pi}\right)^2[L^2 \left(39.0147\, -1.0836 n_f\right)+L \left(1.03408 n_f-95.7303\right)-12.6327 n_f+212.735] \nonumber\\
&&C^{\overline{\rm MS}(2)}_{6}(\alpha_{s},z^2 \mu^2)=\left(\frac{\alpha_s}{2\pi}\right)^2[L^2 \left(42.6881\, -1.14841 n_f\right)+L \left(1.43362 n_f-120.364\right)-13.8654 n_f+254.226] \nonumber\\
&&C^{\overline{\rm MS}(2)}_{7}(\alpha_{s},z^2 \mu^2)=\left(\frac{\alpha_s}{2\pi}\right)^2[L^2 \left(46.01\, -1.20485 n_f\right)+L \left(1.80568 n_f-143.912\right)-15.0307 n_f+295.852] \nonumber\\
&&C^{\overline{\rm MS}(2)}_{8}(\alpha_{s},z^2 \mu^2)=\left(\frac{\alpha_s}{2\pi}\right)^2[L^2 \left(49.0487\, -1.25485 n_f\right)+L \left(2.15437 n_f-166.474\right)-16.1389 n_f+337.394] \nonumber\\
&&C^{\overline{\rm MS}(2)}_{9}(\alpha_{s},z^2 \mu^2)=\left(\frac{\alpha_s}{2\pi}\right)^2[L^2 \left(51.8538\, -1.29974 n_f\right)+L \left(2.48291 n_f-188.142\right)-17.1981 n_f+378.698] \nonumber\\
&&C^{\overline{\rm MS}(2)}_{10}(\alpha_{s},z^2 \mu^2)=\left(\frac{\alpha_s}{2\pi}\right)^2[L^2 \left(54.4622\, -1.34048 n_f\right)+L \left(2.79386 n_f-208.996\right)-18.2144 n_f+419.684] \nonumber\\
&&C^{\overline{\rm MS}(2)}_{11}(\alpha_{s},z^2 \mu^2)=\left(\frac{\alpha_s}{2\pi}\right)^2[L^2 \left(56.9024\, -1.37778 n_f\right)+L \left(3.08931 n_f-229.108\right)-19.1925 n_f+460.29] \nonumber\\
&&C^{\overline{\rm MS}(2)}_{12}(\alpha_{s},z^2 \mu^2)=\left(\frac{\alpha_s}{2\pi}\right)^2[L^2 \left(59.197\, -1.41217 n_f\right)+L \left(3.37098 n_f-248.541\right)-20.1365 n_f+500.487],
\eea
where $L=\ln \left(\frac{z^{2}\mu^{2}e^{2\gamma_E}}{4}\right)$.

The anomalous dimension $\gamma^{\overline{\rm MS}}_{N}$ describes the scale $\mu$ dependence of Wilson coefficients:
\bea
\mu \frac{d C^{\overline{\rm MS}}_{N}(\alpha_{s},z^2 \mu^2)}{d \mu} = \gamma^{\overline{\rm MS}}_{N}(\alpha_{s}) C^{\overline{\rm MS}}_{N}(\alpha_{s},z^2 \mu^2).
\eea
Doing the derivative with respect to $\mu$ on both sides of Eq.~\eqref{eq:OPE}, we achieve the following identity:
\bea\label{eq:ADWilcoe}
\gamma^{\overline{\rm MS}}_{N} = \gamma^{\overline{\rm MS}} + \gamma^{\rm ratio}_{N},
\eea
where $\gamma^{\overline{\rm MS}}$ is the heavy-light quark current anomalous dimension, which satisfies:
\bea
\mu \frac{d \tilde{h}^{\overline{\rm MS}}(z, \mu, P_{z})}{d \mu} = \gamma^{\overline{\rm MS}}(\alpha_{s}) \tilde{h}^{\overline{\rm MS}}(z, \mu, P_{z}),
\eea
and is calculated up to NNLO in literature~\cite{Politzer:1988wp,Ji:1991pr,Braun:2020ymy}:
\bea\label{ADHLC}
&&\gamma^{\overline{\rm MS}}=\frac{\alpha_{s}}{2 \pi}3 C_{F} + \left(\frac{\alpha_{s}}{2 \pi}\right)^2\left(\left(\frac{49}{12}-\frac{\pi ^2}{3}\right) C_A C_F-\frac{5}{3} C_F n_f T_F-\left(\frac{5}{4}-\frac{4 \pi ^2}{3}\right)C_F^2 \right) \nonumber\\
&&+\left(\frac{\alpha_{s}}{2 \pi}\right)^3\left(-\frac{35 n_f^2}{81}+\left(-\frac{332 \zeta (3)}{27}-\frac{172}{81}-\frac{196 \pi ^2}{243}\right) n_f-\frac{178 \zeta (3)}{9}-\frac{61}{12}+\frac{686 \pi ^2}{81}+\frac{380 \pi ^4}{243}\right).
\eea
$\gamma^{\rm ratio}_{N}$ is the anomalous dimension for DGLAP evolution (we use the non-singlet and valence case) which satisfies:
\bea
\mu \frac{d \langle x^N \rangle(\mu)}{d \mu} = -\gamma^{\rm ratio}_{N}(\alpha_{s}) \langle x^N \rangle(\mu).
\eea
Analytical forms of $\gamma^{\rm ratio}_{N}$ up to NNLO are presented in~\cite{Moch:2004pa}. Useful numerical results can be found here~\cite{Retey:2000nq}.

\section{Inverse LaMET Matching}\label{app:inversematching}

In this section, we derive a series expansion formalism to invert the matching kernel. Although we don't use this formalism in our numerical calculation (we invert discrete matrix instead), it can help us understand the log term. 

We rewrite the matching kernel as 
$$
C\left(\frac{x}{y},\frac{\mu}{|x|P_{z}}\right)=\delta\left(1-\frac{x}{y}\right) + \delta C\left(\frac{x}{y},\frac{\mu}{|x|P_{z}}\right),
$$ 
and put it into Eq.~\eqref{eq:matching},
\bea
\tilde{f}(x, P_{z}) = f(x, \mu) + \int_{-1}^{1} \frac{dy}{\left| y \right|} \delta C\left(\frac{x}{y},\frac{\mu}{|x|P_{z}}\right) f(y, \mu) + {\cal O}\left[\frac{\Lambda_{\rm QCD}^2}{x^2 P_{z}^2}, \frac{\Lambda_{\rm QCD}^2}{(1-x)^2 P_{z}^2}\right].
\eea
Rewrite this equation as
\bea\label{eq:inver2}
f(x, \mu) = \tilde{f}(x, P_{z}) - \int_{-1}^{1} \frac{dy}{\left| y \right|} \delta C\left(\frac{x}{y},\frac{\mu}{|x|P_{z}}\right) f(y, \mu) + {\cal O}\left[\frac{\Lambda_{\rm QCD}^2}{x^2 P_{z}^2}, \frac{\Lambda_{\rm QCD}^2}{(1-x)^2 P_{z}^2}\right].
\eea
The $f(y, \mu)$ on the right hand side of Eq.~\eqref{eq:inver2} can been substituted by Eq.~\eqref{eq:inver2} itself iteratively to arrive at Eq.~\eqref{eq:inver3}:
\bea\label{eq:inver3}
&&f(x, \mu) = \tilde{f}(x, P_{z}) - \int_{-1}^{1} \frac{dy}{\left| y \right|} \delta C\left(\frac{x}{y},\frac{\mu}{|x|P_{z}}\right) \tilde{f}(y, P_{z}) 
+ \int_{-1}^{1} \frac{dy}{\left| y \right|} \delta C\left(\frac{x}{y},\frac{\mu}{|x|P_{z}}\right) \int_{-1}^{1} \frac{dw}{\left| w \right|} \delta C\left(\frac{y}{w},\frac{\mu}{|y|P_{z}}\right) f(w, P_{z}) \nonumber\\
&&= \tilde{f}(x, P_{z}) - \int_{-1}^{1} \frac{dy}{\left| y \right|} \delta C\left(\frac{x}{y},\frac{\mu}{|x|P_{z}}\right) \tilde{f}(y, P_{z}) 
+ \int_{-1}^{1} \frac{dy}{\left| y \right|} \delta C\left(\frac{x}{y},\frac{\mu}{|x|P_{z}}\right) \int_{-1}^{1} \frac{dw}{\left| w \right|} \delta C\left(\frac{y}{w},\frac{\mu}{|y|P_{z}}\right) \tilde{f}(w, P_{z}) \nonumber\\
&&-\int_{-1}^{1} \frac{dy}{\left| y \right|} \delta C\left(\frac{x}{y},\frac{\mu}{|x|P_{z}}\right) \int_{-1}^{1} \frac{dw}{\left| w \right|} \delta C\left(\frac{y}{w},\frac{\mu}{|y|P_{z}}\right) \int_{-1}^{1} \frac{dv}{\left| v \right|} \delta C\left(\frac{w}{v},\frac{\mu}{|w|P_{z}}\right) \tilde{f}(v, P_{z}) + ... + {\cal O}\left[\frac{\Lambda_{\rm QCD}^2}{x^2 P_{z}^2}, \frac{\Lambda_{\rm QCD}^2}{(1-x)^2 P_{z}^2}\right]
\eea
Let's keep the series up to NLO, which works for the first non-trivial $\alpha_s$ order of $\delta C$:
\bea
f(x, \mu) 
=&& \tilde{f}(x, P_{z}) - \int_{-1}^{1} \frac{dy}{\left| y \right|} \delta C\left(\frac{x}{y},\frac{\mu}{|x|P_{z}}\right) \tilde{f}(y, P_{z}) + {\cal O}\left[\frac{\Lambda_{\rm QCD}^2}{x^2 P_{z}^2}, \frac{\Lambda_{\rm QCD}^2}{(1-x)^2 P_{z}^2}\right],
\eea
where there is a log term $\sim \ln \left(\frac{\mu^2}{4 x^2 P_z^2} \right)$ in $\delta C\left(\frac{x}{y},\frac{\mu}{|x|P_{z}}\right)$ and $x$ is the momentum fraction of light cone parton.

\section{A Method Solving RG Equation}\label{app:exactscalemethod}

We provide an exact method to DGLAP evolve $f(x,2 x P_z c')$ to scale $\mu$. The exact method makes use of one property of DGLAP: only higher momentum partons can split into lower momentum partons so we only need partons with momentum fraction no less than $x$ to evalutate the DGLAP evolution results for the parton at momentum fraction $x$. Imagine we have each parton evaluated at its own intrinsic scale at the beginning. Note that the largest momentum parton ($x=1$) can evolve with itself. So one can start from the largest momentum parton and evolve it to the intrinsic scale of the second largest momentum parton. Now these two partons are at the same scale so we can evolve them together to the intrinsic scale of the third largest momentum parton. With the same strategy, a recursive operation will finally evolve all the partons to the same scale.

The numerical implementation is the following:\\ 
1) We discretize the momentum region $x \in [0,1]$ into $N_x$ slices, $x_{i}=\frac{i}{N_x}, i=1,..,N_x$;\\
2) We start from $x_{N_x}=1$. We match the quasi-PDF with fixed order matching kernel (Eq.~\eqref{eq:matching}) to scale $2 x_{N_x} P_z c'$ on the light cone (the scale in the $\alpha_s$ is also $2 x_{N_x} P_z c'$):
\bea
f(x,2 x_{N_x} P_z c') = C^{-1}\left(\frac{x}{y}, \frac{2 x_{N_x} P_z c'}{|x| P_{z}}\right) \otimes \tilde{f}(y,P_z).
\eea
Then we evolve $f(x,2 x_{N_x} P_z c')$ to scale $\mu$ and $2 x_{N_x-1} P_z c'$ separately based on DGLAP equation Eq.~\eqref{eq:DGLAP}:
\bea
&&f_{N_x}(x,\mu) = M\Big\{e^{\int_{2 x_{N_x} P_z c'}^{\mu} 
\frac{d\mu'}{\mu'} P[w,\alpha_{s}(\mu')]}\Big\} \otimes f\left(\frac{x}{w},2 x_{N_x} P_z c'\right) \nonumber\\
&&g_{N_x}(x,2 x_{N_x-1} P_z c')= M\Big\{e^{\int_{2 x_{N_x} P_z c'}^{2 x_{N_x-1} P_z c'} 
\frac{d\mu'}{\mu'} P[w,\alpha_{s}(\mu')]}\Big\} \otimes f\left(\frac{x}{w},2 x_{N_x} P_z c'\right), 
\eea
where $M$ is the scale ordering operator. There is no scale mismatch effect near $x=1$;\\
3) For $x_i<x_{N_x}$, we match the quasi-PDF with fixed order matching kernel to scale $2 x_{i} P_z c'$ on the light cone (the scale in the $\alpha_s$ is also $2 x_{i} P_z c'$): 
\bea
f(x,2 x_{i} P_z c') = C^{-1}\left(\frac{x}{y}, \frac{2 x_{i} P_z c'}{|x| P_{z}}\right) \otimes \tilde{f}(y,P_z). 
\eea
Then we reconstruct a PDF as
\bea
f_r(x,2 x_{i} P_z c') = 
\begin{cases}
f(x,2 x_{i} P_z c') & x < x_{i}+\frac{1}{2N_x} \\
g_{i+1}(x,2 x_{i} P_z c') & x > x_{i}+\frac{1}{2N_x} \\
\end{cases},
\eea
where each point at $x>x_{i}-\frac{1}{2N_x}$ is evaluated at its intrinsic physical scale or evolved from its intrinsic physical scale.
Then we evolve $f_r(x,2 x_{i} P_z c')$ to scale $\mu$ and $2 x_{i-1} P_z c'$ separately:
\bea
&&f_{i}(x,\mu) = M\Big\{e^{\int_{2 x_{i} P_z c'}^{\mu} 
\frac{d\mu'}{\mu'} P[w,\alpha_{s}(\mu')]}\Big\} \otimes f_r\left(\frac{x}{w},2 x_{i} P_z c'\right) \nonumber\\
&&g_{i}(x,2 x_{i-1} P_z c')= M\Big\{e^{\int_{2 x_{i} P_z c'}^{2 x_{i-1} P_z c'} 
\frac{d\mu'}{\mu'} P[w,\alpha_{s}(\mu')]}\Big\} \otimes f_r\left(\frac{x}{w},2 x_{i} P_z c'\right). 
\eea
So one can solve from $i=N_x-1,N_x-2...$ to $i=1$;\\
4) Our final PDF is $f(x_i,\mu)=f_{i}(x_i,\mu), i=1,..,N_x$.

\twocolumngrid
\bibliographystyle{apsrev4-1}
\bibliography{ref}

\end{document}